\documentclass[aos]{imsart}

\RequirePackage{amsthm,amsmath,amsfonts,amssymb}
\RequirePackage[authoryear]{natbib}
\usepackage[T1]{fontenc}
\usepackage[utf8]{inputenc}
\usepackage[english]{babel}
\usepackage[font=small, labelfont = bf]{caption}
\usepackage{graphicx}
\usepackage{algorithm}
\usepackage{algorithmic}
\usepackage {graphicx,color}
\usepackage[dvipsnames]{xcolor}
\usepackage{capt-of}
\usepackage{booktabs}
\usepackage{varwidth}
\usepackage{comment}
\usepackage{lineno}
\usepackage{amsbsy}
\usepackage{bm}
\usepackage{textcomp}
\usepackage{amsthm}
\usepackage{hyperref}
\usepackage{blindtext}
\usepackage{booktabs}
\usepackage{placeins}

\usepackage{multirow}
\usepackage{makecell}

\startlocaldefs

\newtheorem{thm}{Theorem}[section]

\newtheorem{condition}{Condition}[section]

\newtheorem{corol}{Corollary}[section]
\newtheorem{remark}{Remark}

\graphicspath{{Images/}}

\newcommand{\N}{\mathbb{N}}
\newcommand{\Z}{\mathbb{Z}}

\newcommand{\R}{\mathbb{R}}
\newcommand{\E}{\mathbb{E}}

\newcommand{\argmin}{\mathop{\rm arg\min}}

\newcommand{\Cov}{\mathop{\rm Cov}}

\newcommand{\IE}{\mathbb{E}}
\newcommand{\IP}{\mathbb{P}}

\makeatletter
\newcommand\code{\bgroup\@makeother\_\@makeother\%\@makeother\~\@makeother\$\@codex}
\def\@codex#1{{\normalfont\ttfamily\hyphenchar\font=-1 #1}\egroup}
\makeatother
\let\proglang=\textsf
\newcommand{\pkg}[1]{{\fontseries{b}\selectfont #1}}

\endlocaldefs

\begin{document}

\begin{frontmatter}
\title{Change point analysis with irregular signals}
\runtitle{Change point analysis with irregular signals}

\begin{aug}
\author[A]{\fnms{Tobias}~\snm{Kley}\ead[label=e1]{tobias.kley@uni-goettingen.de}},
\author[B]{\fnms{Yuhan Philip}~\snm{Liu}\ead[label=e2]{yuhanphilipliu@uchicago.edu}}
\author[C]{\fnms{Hongyuan}~\snm{Cao}\ead[label=e3]{hcao@fsu.edu}}
\and
\author[B]{\fnms{Wei Biao}~\snm{Wu}\ead[label=e4]{wbwu@uchicago.edu}}
\address[A]{Institute for Mathematical Stochastics, Georg-August-Universit\"at G\"ottingen, Germany\printead[presep={,\ }]{e1}}

\address[B]{Department of Statistics, University of Chicago, Chicago, IL, 60637\printead[presep={,\ }]{e2,e4}}
\address[C]{Department of Statistics, Florida State University, Tallahassee, FL, 32306\printead[presep={,\ }]{e3}}
\end{aug}

\begin{abstract}
This paper considers the problem of testing and estimation of change point where signals after the change point can be highly irregular, which departs from the existing literature that assumes signals after the change point to be piece-wise constant or vary smoothly. A two-step approach is proposed to effectively estimate the location of the change point. The first step consists of a preliminary estimation of the change point that allows us to obtain unknown parameters for the second step. In the second step we use a new procedure to determine the position of the change point. We show that, under suitable conditions, the desirable $\mathcal{O}_{\IP}(1)$ rate of convergence of the estimated change point can be obtained. We apply our method to analyze the Baidu search index of COVID-19 related symptoms and find 8~December 2019 to be the starting date of the COVID-19 pandemic.
\end{abstract}



\end{frontmatter}


\section{Introduction}
\label{sec:intro}

Change point detection and localization are classic and reviving topics in many dynamically evolving systems, where a sequence of measurements are recorded and we are interested in determining whether and at what time or location some aspect of the data, such as mean, variance or distribution, changes \citep{Page1955, Page1957}. This problem is of interest in many fields, such as economics, climatology, engineering, genomics, to name just a few. The last few decades witnessed enormous development on this topic from different perspectives including testing the existence of change points and the estimation of their locations. We refer to \cite{csorgo1997limit, aue2013structural, jandhyala2013inference, niu2016multiple} for reviews and recent developments on this topic.  

An important problem in the detection of structural breaks is the detection of mean changes. The simplest case where there is at most one change point has been studied extensively. The first step is to test whether there is any change point. If we reject the null hypothesis that there is no change point, the next step is to make inference on the location of the change point \citep{hawkins1977testing}. The latter problem is nontrivial even for the normal and homoskedastic model or the one-parameter exponential family \citep{sen1975tests, hinkley1970inference, worsley1986confidence, siegmund1988confidence}. 
Recently, the problem of detecting multiple change points has drawn a lot attention \citep{frick2014multiscale, fryzlewicz2014wild, fryzlewicz2018tail, baranowski2019narrowest}.
In particular, functions in the popular \proglang{R} package \pkg{changepoint} achieve linear computational cost when the number of change points increases with the number of observations \citep{KillickFearnheadEckley2012,changepointJSS}. Here it is assumed that, under the alternative, the mean function is piecewise constant. However, in certain applications, it is more plausible to assume that functions between a finite number of change points vary smoothly and/or the error process is dependent. Relevant statistical methods and theory can be found in \cite{Mueller1992, HorvathKokoszka, MallikEtAl2011, MallikEtAl2013, VogtDette2015, dette2020multiscale, bucher2021deviations}.  

\begin{figure}[t]
	\centering
	\includegraphics[width=\textwidth]{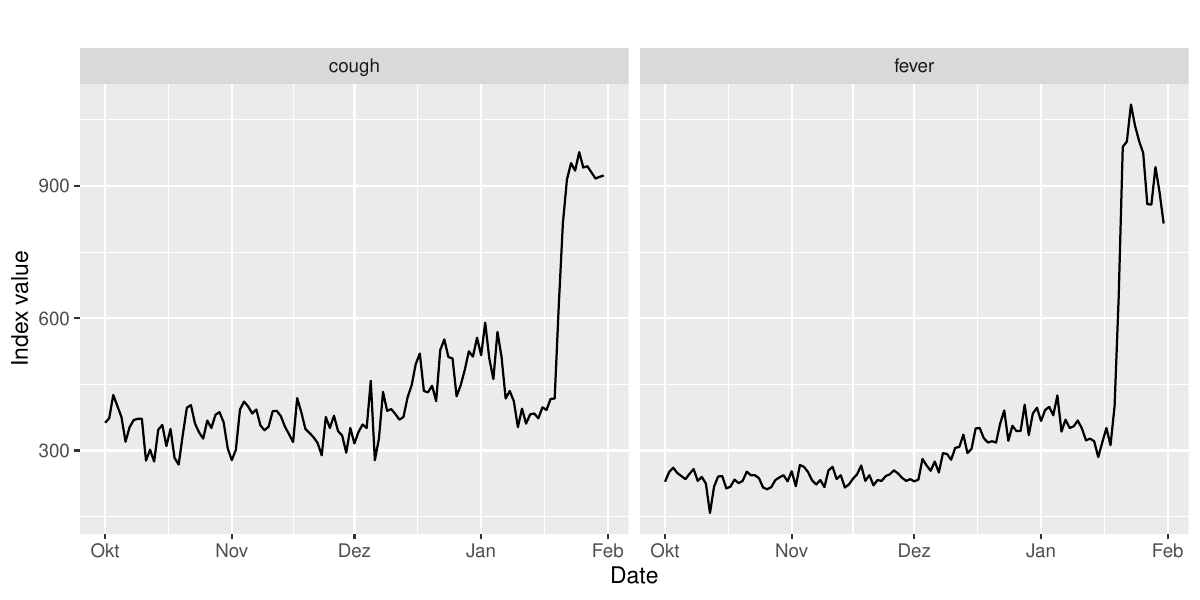} 
	\caption{Daily Baidu search index values for keywords of COVID-related symptoms. Dates are from 1 October 2019 to 31 January 2020. Keywords ``cough'' and ``fever'' are shown in left and right panel, respectively.}
	\label{Fig:covid_for_intro}   
\end{figure}

Unlike existing literature in the previous paragraph, we motivate our research from the fact that signals can be highly irregular after the change point in certain applications. Such irregular signals depart sharply from the constant mean or smoothly varying functions under the alternative--they can vary abruptly. A typical data example with irregular signals is depicted in Figure \ref{Fig:covid_for_intro}. In this dataset, the total number of searches for COVID-19 related symptoms such as ``fever'' through Baidu (the most used search engine in China) is recorded in Hubei Province from 1 October 2019 to 31 January 2020. An extremely important problem in epidemiology is to identify the starting date of the pandemic. However the latter problem is very difficult due to lack of access to the data, and researchers have different results on this; see for example \cite{worobey2021dissecting}, \cite{HuangEtAl2020}, \cite{HuangEtAl2020} and \cite{CDCCovid} among others. In this paper we shall address this important and fundamental problem by using the indirect Baidu search data. In particular we are interested in inferring the start of the COVID-19 pandemic through changes of the Baidu search index by imposing the change point paradigm (\ref{eq:H0}) and (\ref{eq:H1}). Based on the nature of the problem, it seems plausible to assume a constant mean before the change point. After the change point, it does not make sense to assume constant mean or smooth trend as the data exhibit a high level of variation and irregularity. As far as we know, no results in the current change point literature can allow such a framework.

In this paper, we consider the testing and estimation of at most one change point where the null assumes a constant mean and signals under the alternative can be quite general. Our method is offline in the sense that we assume data are fully observed which is different from online change point detection where information accrue over time and we only have data available that has arrived before the current time.  
Differently from \cite{CaoWu2015}, where the interest lies in the multiple testing with clustered signals, we propose new test statistics and a two-step method for detection of irregular signals after the change point. We use a CUSUM-type of statistic to test the global null hypothesis that there is no change point. If this global null is rejected, we develop a two-step method to locate the change point. In the first step, we use the minimum of the batched means as a rough estimation of the change point location. Intuitively speaking, as data before the change point have a constant mean, this minimum falls between the time origin and the true change point. The batched mean effectively smooths out the data and increases the signal-to-noise noise ratio. Equipped with the preliminary estimation from the first step, we are able to estimate the constant mean before the change point and the minimum distance between signals and the constant mean. This allows us to construct a new test statistic to get a refined estimation of the change point in the second step. We get, under suitable conditions, an $\mathcal{O}_{\IP}(1)$ rate of convergence of the estimated change point to the true change point, which fundamentally improves the results in \cite{cao2022testing}, where multiple sequences are needed to estimate the variance due to heteroscedasticity. 

The rest of the paper is organized as follows. In Section~\ref{sec:methodology}, we introduce our main methodology for the global test and the two-step method to locate the change point. Theoretical results are developed in Section~\ref{sec:theory}, where we show Gaussian approximation to the test statistic and the desirable $\mathcal{O}_{\IP}(1)$ rate of convergence of the estimated change point to the true change point. In Section~\ref{sec:simulation}, we investigate the finite sample performance of the new method through simulations. In Section~\ref{sec:real_data}, we apply the new change point estimation method to two datasets: the Baidu search index for COVID-19 related symptoms ``fever'' and ``cough'' in 2019--2020.
All proofs and technical details are relegated to the Appendix (in the Supplementary Material).

\section{Methodology}
\label{sec:methodology}

\subsection{Model for the data and the problems considered}
\label{sec:model_problem}

Suppose we are given noisy data of the form
\begin{equation}\label{eq:model}
X_{t} = \mu_{t} + Z_t, \quad t = 1, \ldots, n,
\end{equation}
where $\mu_{t}$ are means or signals and $(Z_t)_{t \in \mathbb{Z}}$ is a stationary process with mean $0$, auto-covariance function $\gamma(k) = \Cov(Z_{t+k}, Z_t)$, and finite long-run variance 
\begin{equation}\label{eq:lrv}
0 < \sigma_{\infty}^2 := \sum_{k=-\infty}^{\infty} \gamma(k) < \infty.
\end{equation}
Consider the following null hypothesis
\begin{equation}
\label{eq:H0}
H_0: \, \mu_1 = \ldots = \mu_n,
\end{equation}
where the signal is constant (i.\,e., all means $\mu_j$ are equal, but not necessarily zero) and the alternative hypothesis
\begin{equation}
\label{eq:H1}
H_1: \ \exists \, \tau \in \{2, \ldots, n\}, d > 0 \, : \, \mu_1 = \ldots = \mu_{\tau-1}, \quad \mu_{\tau}, \ldots, \mu_n \geq \mu_1 + d,
\end{equation}
where the signal is constant for the first $\tau-1$ observations and the means from the $\tau$th observation onward may then vary arbitrarily as long as they are larger by at least $d$.
We focus on the one-sided case, because such upward shifts to a higher, but non-constant, level are frequently encountered in practice; cf. Figure~\ref{Fig:covid_for_intro}. Yet, to our knowledge, no method that is tailored to this important situation is available to date.
Note that this setting includes the case where $\mu_{\tau} = \ldots = \mu_n \geq \mu_1 + d$, which is to be detected by many traditional methods.
The case of multiple changes (as long as $\mu_j \geq \mu_1 + d$, $j \geq \tau$) is also covered. Then $\tau$ corresponds to the time of the earliest change. But, paradigm~\eqref{eq:H1} goes far beyond these specific cases.
In fact, apart from the one-sidedness of the change in means, we do not require any structure and the signal after the change may be arbitrarily wild.

Given the observations \(X_1, \ldots, X_n\) we aim to develop\enlargethispage{0.5cm}

\begin{itemize}
	\item a hypothesis test to decide whether $H_0$ holds or $H_1$ holds (see Section~\ref{sec:test}), and
	\item a procedure that, under $H_1$, will estimate $\tau$ (see Section~\ref{sec:loc_algo}).
\end{itemize}

Note that, \cite{dette2019detecting, heinrichs2021distribution, VogtDette2015, bucher2021deviations} considered the problem of detecting changes in a sequence of means. Smoothness assumptions are needed for their methods to work. Different from these works, as mentioned before, our methodology does not require such smoothness assumptions.

\subsection{Testing procedure}
\label{sec:test}

Now, we propose a test to distinguish between $H_0$ and $H_1$, defined in~\eqref{eq:H0} and~\eqref{eq:H1}, respectively, when we have $X_1, \ldots, X_n$ that follow~\eqref{eq:model} available.
We use the following quantity:
\begin{equation}
\label{eq:def_T}
\hat T := \min_{j = 1,2,\ldots,n} \sum_{i=1}^j (X_i - \bar{X}_n) / (\sqrt{n} \hat\sigma_{\infty}), \quad \text{where } \bar X_n := \frac{1}{n} \sum_{i=1}^n X_i,
\end{equation}
and $\hat\sigma_\infty^2$ is a consistent estimator for the long-run variance $\sigma_\infty^2$, defined in~\eqref{eq:lrv}.
In Section~\ref{sec:est_lrv} we propose an estimator for $\sigma_{\infty}^2$ that is consistent both under $H_0$ and $H_1$.
In Section~\ref{sec:theory:test} we will show that, under $H_0$, the distribution of $\hat T$ is close, asymptotically, to the distribution of a minimum obtained from a standard Brownian bridge, the quantiles of which can be obtained via simulation or approximated asymptotically.

We will further show that, under $H_{1}$, we have $\hat T \xrightarrow{P} -\infty$, as $n \rightarrow \infty$.
Therefore, we will use $\{\hat T < c\}$ as the rejection area, where $c$ can be chosen as the $\alpha$-quantile of a minimum of a standard Brownian bridge; cf.\ Section~\ref{sec:theory:test}.

\subsection{A two-step locating algorithm}
\label{sec:loc_algo}

\subsubsection{Blocking and estimation of the long-run variance}
\label{sec:blocking}

To reduce the noise from the data and to focus our attention on the signal, we will, here and in the following sections, split the data set into $m := \lfloor n/k \rfloor$ blocks of size $k$ where $k \to \infty$ and $k/n \to 0$. Then we calculate the blocks' sample means as follows:
\begin{equation}
\label{eq:def_R}
R_{j} := \frac{1}{k} \sum_{i = (j-1)k +1}^{jk} X_i, \quad j = 1,2,\ldots, m.
\end{equation}
Our theoretical results and remarks provide guidance on the choice of $k$; cf. Section~\ref{sec:theory:loc_algo}.

From the $R_{j}$ we then obtain
\begin{equation}
\label{eq:def_L}
\hat{L} := \argmin_{i=1, \ldots, m } R_{i}, \quad \hat\ell := k \hat L,
\end{equation}
where $\hat L$ indicates an index of a block likely to have all observations in it prior to the change point and $\hat\ell$ points to the last observation in the $\hat L$th block.
The observations $X_1, \ldots, X_{\hat\ell}$ are approximately stationary and can be used to obtain an estimate for the long-run variance by computing any consistent estimator for the long-run variance from them.
We provide an estimate in Section~\ref{sec:est_lrv} and an asymptotic theory in Section~\ref{sec:theory:loc_algo}.
More generally than~\eqref{eq:def_L}, we can let $\hat L := \max\{i : R_i \leq R^{(J)}_m\}$, for fixed $J$, where $R^{(J)}_m$ denotes the $J$th smallest value amongst the block averages $R_1, \ldots, R_m$. With this generalized definition we have more data to estimate the long-run variance.

\subsubsection{Estimate for long-run variance}
\label{sec:est_lrv}

Having blocked the data as delineated in Section~\ref{sec:blocking}, we derive the index $\hat{\ell}$, which satisfies $\hat\ell < \tau$ with high probability.
To estimate the long-run variance, we suggest
\begin{equation}
\label{eq:lrv_est_mu0hat}
\hat{\sigma}^2 := \frac{k}{\hat{\ell} - k + 1} \sum_{s=k}^{\hat{\ell}} \Big( R_{s/k} - \hat\mu_0 \Big)^2, \quad \text{where} \quad R_{s/k} := \frac{1}{k} \sum_{i = s - k +1}^{s} X_i, \quad  \hat\mu_0 := \frac{1}{ \hat\ell } \sum_{i=1}^{\hat\ell} X_{i},
\end{equation}
where the definition of the overlapping block averages $R_{s/k}$ extends the one for the non-overlapping blocks in~\eqref{eq:def_R}, and $\hat\mu_0$ is a preliminary estimate for $\mu_1$.
We use overlapping blocks in this section, as this has been shown to reduce the asymptotic mean squared error; cf. \cite{Lahiri1999}.
The estimate for the long-run variance is motivated by the fact that
\[\IE[ ( \sqrt{k} (R_{s/k} - \IE R_{s/k}) )^2 ] \to \sigma_{\infty}^2, \quad s = 1, \ldots, \tau - 1  \]
together with $\tau > \hat \ell \to \infty$, with high probability, cf.\ Lemma~A.1, and $\hat\mu_0$ being a consistent estimator for $\mu_1$, cf.\ Lemma~A.3.
Lemmas~A.1 and~A.3, such as all other references of the form A.x are in the Supplementary Material.
Estimates of a similar nature were previously considered, for example, by~\cite{MiesSteland2023}, \cite{Zhou2013}, and \cite{PeligradShao1995}.
The important novelty of the estimate proposed in~\eqref{eq:lrv_est_mu0hat} lies in the data-dependent segment selection via the index $\hat\ell$. This ensures consistency under $H_0$ and $H_1$, but makes the rigorous analysis more challenging. At a technical level, proving consistency requires a maximal inequality for quadratic forms.

\subsubsection{Locating algorithm: step 1}
\label{sec:loc_algo_step1}

The aim of step 1 is to obtain an improved estimate for $\mu_1$ and an estimate towards $d$, which we describe as follows.
We block the data as described in Section~\ref{sec:blocking} and obtain the block-wise averages $R_{j}$, the index $\hat L$ that indicates a block with data from before the change, the index $\hat\ell$ that points to the last observation of the $\hat L$'s block, and the preliminary estimate $\hat\mu_0$, defined in~\eqref{eq:lrv_est_mu0hat}, for $\mu_1$.
We then compute `test statistics' $\hat D_j$ and `test decisions' $\hat I_j$ as
\begin{equation}\label{def:Dj}
\hat D_j := \sqrt{k} (R_{j} - \hat{\mu}_0 ) / \hat\sigma_{\infty} \quad \text{and} \quad
\hat I_j = \begin{cases}
1 & \text{if $\hat D_j \geq z_{1 - 1 / m}$} \\
0 & \text{otherwise.}
\end{cases}
\end{equation}
The long-run variance estimate $\hat\sigma_{\infty}^2$ can be specified by the user. Our estimator
$\hat{\sigma}^2$, defined in~\eqref{eq:lrv_est_mu0hat}, is a canonical choice, but any consistent estimate can be used.
The quantity $z_{\alpha}$, $\alpha \in (0,1)$, denotes the $\alpha$th-quantile of the standard normal distribution, and $m := \lfloor n/k \rfloor$, as before.
We then compute
\begin{equation}
\label{def_eta}
\hat\eta := \argmin_{t=1,\ldots,m -1}  \sum_{j=1}^{m}\left(\hat I_{j}-1_{[t+1, m]}(j)\right)^{2} = \argmin_{t=1,\ldots,m -1} \left[\sum_{j=1}^{t} \hat I_{j}+\sum_{j=t+1}^{m}\left(1-\hat I_{j}\right)\right].
\end{equation}
Finally, we obtain the preliminary estimates for $\mu_1$ and towards $d$ by
\begin{equation}
\label{def_mu1hat_dhat}
\hat{\mu}_1  := \frac{1}{k \hat{\eta}} \sum_{i = 1}^{k \hat{\eta}} X_i, \quad \hat{d} : = \min_{\substack{i =  k (\hat\eta + 1) + 1,\\ \ldots, n-k+1}} \ \frac{1}{k} \sum_{j = i}^{i+ k-1} (X_j - \hat{\mu}_1).
\end{equation}

Some comments on the motivation for these estimates are in order. 
Denote by $\eta := \lfloor \tau/k \rfloor$ the index of the last block for which the signal is still constant; i.\,e., $\eta k + 1 \leq \tau \leq (\eta+1) k$, where $\tau$ is the index of the change; cf. the alternative hypothesis~\eqref{eq:H1} considered.
Then we have $\mathbb{E} R_{j} = \mu_1$ for $j=1,\ldots,\eta$ and $\mathbb{E} R_{j} > \mu_1$ for $j=\eta+1, \ldots, m$.
The intuition behind the estimate $\hat{\mu}_0$ is as follows.
Since $R_{1}, \ldots, R_{\eta}$ fluctuate about their common $\mu_1$, but $R_{\eta+1}, \ldots, R_{m}$ fluctuate about means that are strictly larger than $\mu_1$, we have that $\hat L/\eta$ is stochastically bounded away from 0 and 1, namely, for every $\varepsilon > 0$ there exists $\delta > 0$ such that $\IP( \hat L / \eta \in [\delta, 1-\delta]) \geq 1 - \varepsilon$; cf. Lemma~A.1.
Thus, we expect to average $k \hat L \asymp k \eta$ of the pre-change observations to obtain $\hat\mu_0$. In particular, if $\tau$ diverges at the same rate as $n$, then $\hat\mu_0$ can be expected to be a $\sqrt{n}$-consistent estimate for $\mu_1$.

\begin{figure}[t]
	\centering
	\includegraphics[width=0.48\textwidth]{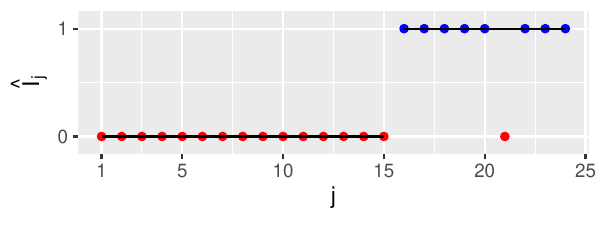}
    \includegraphics[width=0.48\textwidth]{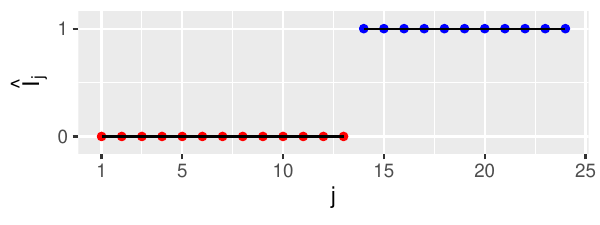}
	\caption{Test decisions $\hat I_j$ for the cough (left) and fever data (right) analysis described in Section~\ref{sec:da:baidu}. Fitted step function $j \mapsto 1_{[\hat\eta+1,m]}(j)$ is indicated by solid black line.}
	\label{Fig:motivation_eta_hat}   
\end{figure}

The rationale behind $\hat\mu_1$ is that by replacing $\hat L$ in $\hat\mu_0$ by $\hat\eta$ we will obtain an improved estimate for $\mu_1$ as we expect the estimate $\hat\eta$ to be closer to $\eta$ than $\hat L$.
Note that, once $\hat\eta$ is available, we use $\hat\eta$ instead of $\hat L$.
The intuition behind $\hat\eta$ is as follows.
The test decisions $\hat I_j$ indicate whether a block is from before the change, where we have $\mathbb{E} R_{j} = \mu_1$, or after the change, where $\mathbb{E} R_{j} > \mu_1$.
Thus, the sequence $\hat I_1, \ldots, \hat I_{m}$ of test decisions is an empirical version of the sequence $I_1, \ldots, I_m$ with $I_j := 0$, $j \leq \eta$ and $I_j := 1$, $j > \eta$, which is known to be a step function.
The estimate $\hat\eta$ is obtained by fitting a step function to the sequence $\hat I_1, \ldots, \hat I_{m}$ of test decisions that jumps from 0 to 1 at the block that includes the change.
Fitting the step function can be seen as a different type of smoothing that we employ to reduce noise in the sequence of test decisions. A graphical illustration of the type of smoothing employed in the estimation of $\eta$ for the empirical example of Section~\ref{sec:da:baidu} is shown in Figure~\ref{Fig:motivation_eta_hat}.
The threshold $z_{\alpha}$, employed in~\eqref{def:Dj}, is chosen with the quantile level $\alpha = 1-1/m$ tending to one to avoid too many false rejections among the blocks prior to the change.

\subsubsection{Locating Algorithm: Step 2}
\label{sec:loc_algo_step2}
In this section we define the novel estimate for the time $\tau$ where the change occurs; cf.~\eqref{eq:H1}.
Consider
\begin{equation}
\label{eq:def_tauhat}
\hat\tau := \arg\min_{j = 2, \ldots, n} \sum_{t = 1}^{j-1} (X_t - \hat{\mu}_1 - \rho \hat{d} ),
\end{equation}
where $\rho \in (0,1)$ is a tuning parameter. In Section~\ref{sec:theory:loc_algo} we provide rigorous theory for $\hat\tau$ which sheds light on the choice of the tuning parameter. As a rule of thumb, we use $\rho = 1/2$. It should be noted that,  in a similar vein, \cite{ChenWangSamworth2021} also uses $1/2$ as a factor for an online changepoint procedure based on likelihood ratio test statistics. 

In Section~\ref{sec:da:baidu}, we study the fundamentally important and much debated problem of inferring the beginning of the COVID-19 pandemic. To this end we employ our estimate $\hat\tau$ to search indices obtained from Baidu. We will see that our method reveals a plausible date, where traditional methods fail completely.

\section{Theory}
\label{sec:theory}


\subsection{Assumptions on the noise process}

To derive meaningful results regarding the statistical properties of our proposed methods, some assumptions regarding the noise process $(Z_t)_{t \in \mathbb{Z}}$ that appears in model~\eqref{eq:model} are in order.

We employ the framework of functional dependence measure introduced in \cite{Wu2005}. In this framework, we view the causal stationary process $(Z_t)_{t \in \Z}$ as outputs from a physical system as follows
\begin{equation}
\label{eq:model_Z}
Z_{t}=G(\ldots, \varepsilon_{t-1}, \varepsilon_{t}),
\end{equation}
where $(\varepsilon_{t})_{t \in \Z}$, i.\,i.\,d., is the input information of this system and $G$ is an $\R$-valued measurable function that can be thought of as a filter or, intuitively, ``mechanism'' of this system. Many widely used, linear and nonlinear time series, including ARCH, threshold autoregressive, random coefficient autoregressive and bilinear autoregressive processes follow the framework of (\ref{eq:model_Z}); see for example \cite{MR1079320, MR991969, wu2011asymptotic} among others. Then with this system, we measure the dependence from how much the outputs of this system will change if we replace the input information at time $t = 0$ with an i.\,i.\,d.\ copy $\varepsilon_{0}'$. Assume $\E |Z_{i}|^{\theta} < \infty$, $\theta \ge 1$. For a single observation at time $i$, we define the functional dependence measure as follows:
\begin{equation}
\label{eq:def_delta_tp}
\delta_{i, \theta}=(\E |Z_{i}-Z_{i,\{0\}}|^{\theta})^{1/\theta},  \quad \text{where} \quad Z_{i,\{0\}} = G(\ldots, \varepsilon_{-1}, \varepsilon_{0}',\varepsilon_{1}, \ldots, \varepsilon_{i} ).
\end{equation}

To measure the temporal dependence for the whole time series, we define the cumulative dependence measure of $\left(Z_{i}\right)_{i \geq n}$ on $\varepsilon_{0}$:

\begin{equation}
\label{eq:def_Theta_tp}
\Theta_{n, \theta}=\sum_{i \geq n} \delta_{i, \theta}, \quad n \geq 0.
\end{equation}

\subsection{Testing procedure}
\label{sec:theory:test}

We now state results on the test described in Section~\ref{sec:test}. The first result provides the asymptotic distribution under the null hypothesis, and the second one asserts asymptotic consistency under $H_1$, where we will allow $d = d_n$ and $\tau = \tau_n$ to depend on $n$ without making this explicit in the notation.

\begin{thm}\label{thm:size_test}
	Assume that the short-range dependence condition holds:
	\begin{equation}\label{eq:srd}
	\Theta_{0, 2}=\sum_{i \geq 0} \delta_{i, 2} < \infty. 
	\end{equation}
	(i) Under $H_0$, we have, as $n \rightarrow \infty$, that
	\begin{equation}
	\sup_{x\le 0} |\mathbb{P}(\hat T \leq x) - e^{-2 x^2}| \to 0. 
	\end{equation}
	(ii) Under $H_1$, assume $(\tau / n) (1-\tau/n) d \sqrt{n} \rightarrow \infty$. Then, we have, as $n \rightarrow \infty$, that
	\[\hat T \to - \infty \quad \textit{in probability}.\]
\end{thm}
The proof is deferred to Section~A.1.
Theorem \ref{thm:size_test}(i) suggests that, given level $\alpha \in (0, 1)$, we can use the $\alpha$-quantile of the limit, $-(-0.5 \log \alpha)^{1/2}$, as the cutoff value to test $H_0$. Let~$\mathbb{B}$ denote a standard Brownian motion and $\mathbb{B}_1(u) = \mathbb{B}(u) - u \mathbb{B}(1)$ be the Brownian bridge. For all $x \le 0$, we have
\begin{equation}\label{eq:rbb}
\mathbb{P}\Big( \inf_{u \in [0, 1]} \mathbb{B}_1(u) \leq x \Big) = e^{-2 x^2};
\end{equation}
cf.\ equation (9.41) in~\cite{Billingsley1999}.
When $n$ is relatively small, a refined approximation of $\mathbb{P}(\hat T \leq x)$ is $\mathbb{P}( T^\circ \leq x )$, where the discretized version $T^\circ = \min_{j \in \{1, 2,\ldots,n \}} \mathbb{B}_1(j/n)$ and its distribution can be obtained by extensive simulation. The test based on the latter can have a more accurate performance. Theorem~\ref{thm:size_test}(ii) implies that for any $q \in \mathbb{R}$, $\mathbb{P}( \hat T \le q) \to 1$.


\subsection{Locating algorithm}
\label{sec:theory:loc_algo}

\enlargethispage{1cm}
To establish a convergence theory for the estimated change points, we will require the following assumption on temporal dependence.

\begin{condition}
	\label{cond:21}
	$(Z_i)_{i \in \Z}$ satisfies that the $\theta$-th moment $H_{\theta} :=  (\E |Z_i|^{\theta})^{1/\theta} < \infty$, where $\theta > 2$. Assume that any one of the following holds
	\begin{itemize}
		\item $\theta>4$ and $\Theta_{n, \theta}=\mathcal{O}\big(n^{-\gamma_{\theta}}(\log n)^{-A}\big)$, as $n \to \infty$, for $A> 2 \left(1 / \theta+1+\gamma_{\theta}\right) / 3$, where
		\[\gamma_{\theta}=(\theta^{2}-4+(\theta-2) \sqrt{\theta^{2}+20 \theta+4})/(8 \theta);\]
		\item $2 < \theta \le 4$ and $\Theta_{n, \theta}=\mathcal{O}\big(n^{-1}(\log n)^{-A}\big)$, as $n \to \infty$, with $A>3 / 2$.
	\end{itemize}
\end{condition}

Condition \ref{cond:21} holds, for example, under the geometric moment contraction $\delta_{n, \theta} = \mathcal{O}(\rho^n)$ for some $\rho \in (0, 1)$, which is satisfied for many nonlinear time series models; see for example, \cite{MR2351105} or \cite{wu2011asymptotic}. In general, it can be weaker since it allows polynomially decaying functional dependence measures. By Corollary 2.1 in \cite{BerkesLiuWu2014}, Condition \ref{cond:21} implies the following optimal Koml\'{o}s–Major–Tusn\'{a}dy result: on a possibly richer probability space $(\Omega_{c}, \mathcal{A}_{c}, \mathrm{P}_{c})$, there exists $(Z_i^c)_{i \in \mathbb Z} \stackrel{\mathcal{D}}{=} (Z_i)_{i \in \mathbb Z} $, and a standard Brownian motion $\mathbb{B}_c (\cdot)$  such that
\begin{equation}
\label{eq:kmtblu}
\sum_{i=1}^n Z_i^c = \sigma_{\infty} \mathbb{B}_c (n) + o_{a.s.}(n^{1/\theta}).
\end{equation}

The next result asserts that $\hat\sigma^2$, defined in~\eqref{eq:lrv_est_mu0hat}, is a consistent estimator for $\sigma_{\infty}^2$. In the statement of the proof we write $a_n \ll b_n$ or $b_n \gg a_n$ to mean that $a_n = o(b_n)$, as $n \to \infty$. The quantities $d = d_n$ and $\tau = \tau_n$ are the ones from~\eqref{eq:H1}, which we allow to depend on $n$ without making this explicit in the notation, and $k$ is the user-chosen block size; cf. Section~\ref{sec:blocking}. Further, $m := \lfloor n/k \rfloor$ and $\eta := \lfloor \tau/k \rfloor$, as before.
\begin{thm}\label{thm:sigmahat}
	Assume that Condition~\ref{cond:21} holds, $d \gg n^{-1/\theta}$ and $n^{2/\theta} \log(n) \ll k \ll \tau$.
	   Then,
	\[ \hat\sigma^2 = \sigma_{\infty}^2 + \mathcal{O}\big( 1/k \big) 
        + \mathcal{O}_{\IP}\big( \eta^{2/\min(4, \theta) -1} \big), \quad \text{as $n \to \infty$}.
	\]
\end{thm}
\noindent
The proof is deferred to Section~A.2.

\begin{remark}
(i) When $\theta \ge 4$, the bound in Theorem \ref{thm:sigmahat} becomes $\mathcal{O}\big( 1/k \big) + \mathcal{O}_{\IP}\big( \eta^{-1/2} \big)$. Intuitively, the $\mathcal{O}( 1/k )$ and $\mathcal{O}_{\IP}( \eta^{-1/2} )$ can be seen as the rate for the bias and a centered version of the estimate $\hat\sigma^2$, respectively. The rate of the bias follows from $\sum_{u=-\infty}^{\infty} |u \gamma(u)| < \infty$, which is satisfied under Condition~\ref{cond:21}; cf.\ Lemma~A.2.

    (ii) The choice of the block size $k$, in Theorem~\ref{thm:sigmahat}, is limited by the rates of $n^{2/\theta} \log(n)$ from below and by $\tau$ from above. The lower bound assures a sufficient noise reduction and is smaller if tails of the noise are lighter. The upper bound implies $\eta \to \infty$.
    Generally, in practice, the moment of the noise $\theta$ is unknown. The nonadaptive block length $k = \lceil n^{1/3}\rceil$ is a simple, yet effective choice that satisfies the condition of Theorems~\ref{thm:sigmahat} and~\ref{thm:tauhat} if $\theta > 6$. \cite{BuhlmannKunsch1999} found that the $n^{1/3}$-choice performs quite well in most cases. 

    (iii) The gap $d$ is allowed to vanish asymptotically for our result, as long as the rate of decay is slower than $n^{-1/\theta}$. The conditions on $d$ and $k$ have to be satisfied for the same $
    \theta$. This means that if $d$ decays slowly and Condition~\ref{cond:21} is satisfied for a large $\theta$, then $k$ can be chosen smaller and the result still holds.

\end{remark}

Our main result of this section is regarding a bound for the error of estimating $\tau$ by $\hat\tau$. We provide this bound in terms of the minimum gap to the signal averaged over sliding blocks. More precisely, defining
\begin{equation}
\label{def_dstar}
d_* : = \min_{\substack{i =  k (\eta + 1) + 1,\\ \ldots, n-k+1}} \ \frac{1}{k} \sum_{j = i}^{i+ k-1} (\mu_j - \mu_1),
\end{equation}
then we have
\begin{thm}
	\label{thm:tauhat}
     Assume Condition~\ref{cond:21}, $d \gg n^{-1/\theta}$, $n^{2/\theta} \log(n) \ll k \ll \tau$, $n-\tau \geq 2 k$, and that there exists a constant $K > \rho$ with $d > K d_*$, where $\rho$ is the tuning parameter from the definition of $\hat\tau$. Let the estimator $\hat\sigma_{\infty}^2$, used in~\eqref{def:Dj}, be consistent for the long-run variance $\sigma_{\infty}^2$; i.\,e., $\hat\sigma_{\infty}^2 = \sigma_{\infty}^2 + o_{\IP}(1)$.
    Then, 
    \[\hat\tau = \tau + \mathcal{O}_{\IP}\big( d_*^{-\theta/(\theta-1)} \big),\]
    as $n \to \infty$.
\end{thm}
\noindent
The proof is deferred to Section~A.3.

\begin{corol}
Under the conditions of Theorem~\ref{thm:tauhat}, we have:\\
	(i) If $d$ is bounded away from zero (i.\,e., if there exists a constant $M $ with $0 < M < d$), then $\hat\tau_n = \tau + \mathcal{O}_{\IP}(1)$, as $n \to \infty$.\\
	(ii) If $d$ is unbounded (i.\,e, $d \to \infty$, as $n \to \infty$), then $\IP(\hat\tau_n = \tau) \to 1$, as $n \to \infty$.
\end{corol}

\begin{remark}

    (i) Under the general alternative~\eqref{eq:H1} we do not impose any regularity conditions, apart from the one sidedness of the change. Hence, it is possible that the gap $d \leq \min_{t = \tau, \ldots, n} (\mu_t - \mu_1)$ is determined by an individual (noisy) observation, which is not enough to consistently estimate $d$ itself. We show (Lemma~A.5) that our proposed $\hat d$ consistently estimates $d_*$. By definition $d_* \geq d$; i.\,e., $d_*$ provides an upper bound for the gap $d$. The regularity condition $d > K d_*$, for a constant $K > \rho$, requires that $d_*$, which is tractable, also facilitates a lower bound with respect to $d$. The regularity condition $d > K d_*$, for a constant $K > \rho$, is sufficient to estimate $\tau$ by $\hat\tau$.
       
    (ii) The following example illustrates a situation where $d > K d_*$ is satisfied, for a constant $K > \rho$: say $\mu_j - \mu_1 = m\big( (j-\tau) / (n-\tau) \big)$ for a function $m: [0,1] \to (0,\infty)$ of bounded total variation $\| m \|_{\rm TV} < \infty$ and let $K := 1 / \big(1 + \| m \|_{\rm TV} / \big(k \inf_{x \in [0,1]} m(x)\big) \big)$. Note that $K > \rho$ if $k > \rho \|m\|_{\rm TV} / \big( (1-\rho) \inf_{x \in [0,1]} m(x) \big)$ and
    \begin{equation*}
        \begin{split}
            d_* 
            & \leq \inf_{x \in [0,1]} m(x) + \frac{1}{k} \| m \|_{\rm TV}
            = K^{-1} \inf_{x \in [0,1]} m(x)
            \leq K^{-1} \min_{t \in \tau, \ldots, n} m\Big( \frac{j-\tau}{n-\tau} \Big) =: K^{-1} d.
        \end{split}
    \end{equation*}
    As an example, take a continuously differentiable function~$f$ and add a finite number of jump discontinuities at distinct $x_1, \ldots, x_b$: i.\,e., $m(x) := f(x) + \sum_{i=1}^b \delta_i I\{x \leq x_i\}$. Then $m$ is of bounded variation: $\| m \|_{\rm TV} = \int_0^1 |f'(t)| {\rm d}t + \sum_{i=1}^b |\delta_i|$.

    (iii) If the true $d$ were known, we could use the following estimate for $\tau$:
    \begin{equation}
        \label{def:tildetau}
        \tilde \tau := \arg\min_{j = 2, \ldots, n} \sum_{t = 1}^{j-1} (X_t - \hat{\mu}_1 - \rho d ).
    \end{equation}
    Following the lines of our proof for Theorem~\ref{thm:tauhat}, it can be shown that $\tilde\tau = \tau + \mathcal{O}_{\IP}\big( d^{-\theta/(\theta-1)} \big)$, for all $\rho \in (0,1)$, without an assumption regarding $d_*$. Note that $\tilde \tau$ is only available if $d$ is known.

    (iv) The conditions regarding $\tau$ allow for `early' and `late' changes. In particular, we do not require that $\tau \asymp n$. The requirement $k \ll \tau$ ensures that there is an increasing number of blocks before the change. The requirement $n-\tau \geq 2 k$ is slightly weaker and ensures that there is at least one complete block after the change. The requirement $n-\tau \geq 2 k$ is needed to estimate $d_*$ (see Lemma~A.5).
\end{remark}

\clearpage

\section{Monte Carlo studies}
\label{sec:simulation}

\subsection{Models considered}
\label{sec:simulation:mdl}

We assess the finite sample performance of both the testing procedure (Section~\ref{sec:test}) and the two-stage locating algorithm (Section~\ref{sec:loc_algo}). Our experiments employ data crafted via the signal plus noise model delineated in~\eqref{eq:model}.

For the noise component, we utilize a threshold AR model~\citep{MR1079320} as follows:
\begin{equation}\label{eqn:defZprime}
Z'_i = \theta\left( |Z'_{i-1}|+ |Z'_{i-2}|\right)+ \varepsilon_{i},
\end{equation}
where \( \theta \) is the parameter governing temporal dependence, and the i.\,i.\,d.\ innovations \( \varepsilon_i \) follow the normal distribution $\mathcal{N}(0, 0.5^2)$. Within this model, a higher absolute value of \( \theta \) indicates stronger temporal dependence. The process remains stationary provided $|\theta| < 0.5$.
The noise process $(Z_i)$ is obtained by centering $Z_i$ as $Z_i := Z_i' - \IE(Z_i')$.
Three digits behind the comma approximations to the values we used are in Table~\ref{tab:sigma}. For $\theta < 0$ we use that the expectation of the process for $\theta$ and $-\theta$ have the same long-run variance and the expectation differs only in sign. Further, for $\varepsilon_i  \sim \mathcal{N}(0,\xi^2)$, $\xi > 0$, we obtain expectation and long-run variances by multiplying the ones from Table~\ref{tab:sigma} with $\xi$ and $\xi^2$, respectively. For example, for $\theta = -0.2$ and $\varepsilon_i \sim \mathcal{N}(0, 0.5^2)$ we use $\IE Z'_i = -0.343 \cdot 0.5$ 
and $\sigma_{\infty}^2 = 1.332 \cdot 0.5^2$.
\begin{table}[t]
\caption{Simulated expectation $\IE Z'_i$ and long-run variance $\sigma_{\infty}^2$ of $(Z'_i)$, defined in~\eqref{eqn:defZprime}, for the case when $\varepsilon_i  \sim \mathcal{N}(0,1)$.}
\label{tab:sigma}
\centering
\begin{tabular}{ccc}
\toprule
$\theta$ & $\IE Z'_i$ & $\sigma_{\infty}^2$ \\
\midrule
0.2 & 0.343 & 1.332 \\
0.3 & 0.577 & 2.104 \\
0.4 & 0.988 & 5.782 \\
\bottomrule
\end{tabular}
\end{table}

Regarding the signal \( \mu_t \), we examine two scenarios: 
(i) under the null hypothesis $H_0$, as defined in~\eqref{eq:H0}, the signal remains constant at $\mu_1 = 0$.
(ii) Under the alternative hypothesis $H_1$, as defined in~\eqref{eq:H1}, the signal generation model is as follows:
\begin{equation}\label{eqn:signal_flux}
\mu_t = 
\begin{cases}
\mu_1 := 0 & \text{for } t = 1, \ldots, \tau-1 \\
\mu_1 + s\left( \frac{2t - 3\tau + \tau'}{\tau' - \tau} \right) & \text{for } t = \tau, \ldots, \tau' \\
\mu_1 + s\left( 2 + \exp\left(\frac{2(t - \tau')}{\tau'' - \tau'}\right) \right) & \text{for } t = \tau' + 1, \ldots, \tau'' \\
\mu_1 + s\left( 2 + \exp(2) \cdot \frac{2n - \tau'' - t}{2n - 2\tau''} \right) & \text{for } t = \tau'' + 1, \ldots, n,
\end{cases}
\end{equation}

In this model, the parameter \( s \) defines the magnitude of deviation from the baseline mean state (\( \mu_1 = 0 \)) for \( t < \tau \) to the varied mean state for \( t \geq \tau \). This model is designed to reflect trends similar to those observed in the search engine index data depicted in Figure~\ref{Fig:covid_for_intro}.

\begin{figure}[t]
	\centering
	\includegraphics[width=\textwidth]{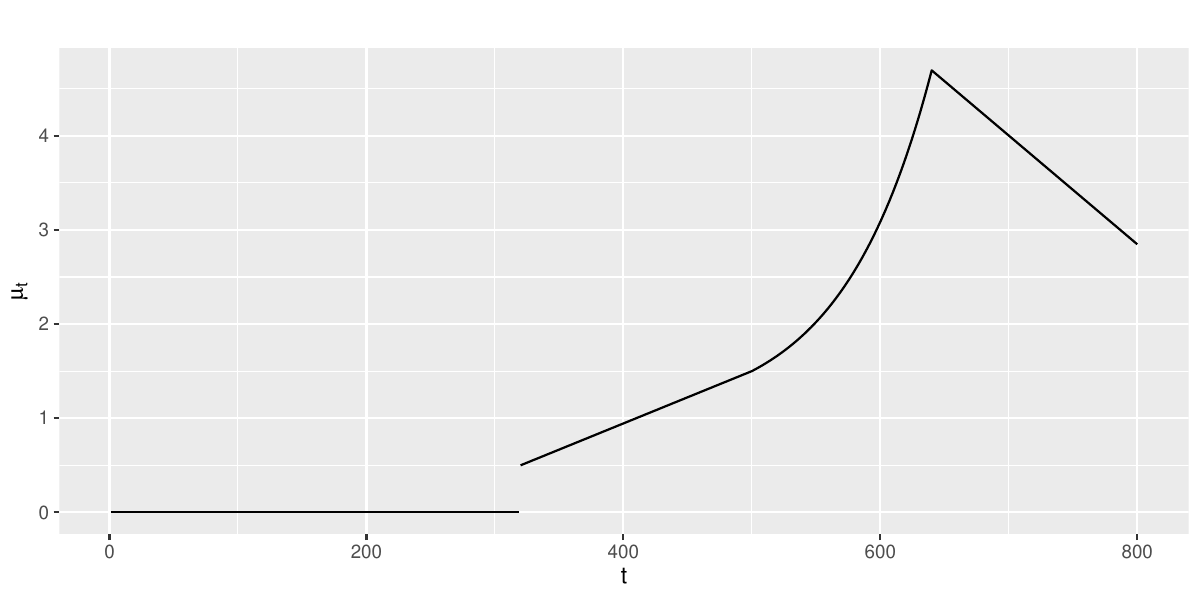}
	\caption{Illustration of $(\mu_t)$ with $n = 800$, $\tau = 320$, $\tau' = 500$,  $\tau'' = 640$,  and $s = 0.5$.}
	\label{Fig:mu_x}
\end{figure}

Figure~\ref{Fig:mu_x} showcases an example of the signal \( (\mu_i) \). It highlights the increase in signal strength after the initial change point at \( \tau = 320 \), where it rises by at least \( s = 0.5 \) above the stable level of \( \mu_1 = 0 \). Beyond the first change point \( \tau \), a second significant change occurs at \( \tau'' = 640 \), where the signal further elevates, reaching at least \( 8s = 4 \) above the initial \( \mu_1 = 0 \) level. This pattern echoes our observations in real-world data.

\subsection{Synthetic data under the null hypothesis}\label{sec:sim:H0}

In this section, we illustrate that the testing procedure described in Section~\ref{sec:test} has the correct size, asymptotically.
We employ data structured as detailed in Section~\ref{sec:simulation:mdl}, operating under a constant signal (i.e., \(H_0\)).

We modulate the sample size, selecting $n$ from \(50, 100, 300, 500, 2000\), and adjust the dependence parameter $\theta$ from $-0.4$, $-0.2$,  $0$ (independence), $0.2$, $0.4$. The significance level remains fixed at $\alpha = 0.05$. We use the true long-run variance $\sigma^2_{\infty}$ instead of $\hat\sigma^2_{\infty}$ in~\eqref{eq:def_T}; cf.\ Table~\ref{tab:sigma}. The empirical sizes, derived from 100,000 replications, are summarized in Table~\ref{tab:ratio_null}.

Analyzing Table~\ref{tab:ratio_null}, it is evident that the rejection ratios—serving as proxies for type-I error—gravitate closer to the target significance level of \(\alpha = 0.05\) as the sample size \(n\) expands and temporal dependence weakens (absolute value of \(\theta\) shrinks). This observation aligns seamlessly with our theoretical framework presented in Section~\ref{sec:theory:test}. On juxtaposing the two methodologies, the finite-sample Gaussian approximation-based testing procedure emerges superior in smaller datasets (\(n = 50, 100, 300, 500\)), compared to its asymptotic counterpart. However, the latter's performance converges with the finite-sample approach as the sample size surges to \(n = 2000\). This implies that, for shorter datasets, the finite-sample Gaussian approximation can be advantageous. Conversely, for longer datasets, the more computationally economical asymptotic approach becomes viable.

Applying the test with the estimated long-run variance leads to higher error rates, in particular for small sample sizes, when it its difficult to estimate the long-run variance. Results are provided in Appendix~B (in the Supplementary Material).

\begin{table}[t]
    \caption{\label{tab:ratio_null}Rejection ratios for change point testing procedure under the null hypothesis; cf. \eqref{eq:H0}.}
    {\footnotesize
        \hspace*{-0.2cm}
        \begin{tabular}{ccccccc} 
            \toprule 
            Approximation & & \multicolumn{4}{c}{$\theta$} \\ 
            \cmidrule{3-7} 
             Method & $n$ & {-0.4} & {-0.2} & {0}  & {0.2}  & {0.4} \\ 
            \midrule 
            Asymptotic 
            &    50 & 1.41\% & 2.70\% & 3.28\% & 3.18\% & 1.63\% \\
            &   100 & 2.26\% & 3.40\% & 3.74\% & 3.69\% & 2.41\% \\
            &   300 & 3.23\% & 3.92\% & 4.21\% & 4.08\% & 3.39\% \\
            &   500 & 3.50\% & 4.17\% & 4.38\% & 4.30\% & 3.56\% \\
            &  2000 & 4.18\% & 4.54\% & 4.54\% & 4.65\% & 4.23\% \\
            \addlinespace
            Finite-sample
            &    50 & 2.10\% & 4.08\% & 5.00\% & 4.79\% & 2.44\% \\
            &   100 & 2.96\% & 4.54\% & 5.00\% & 4.90\% & 3.14\% \\
            &   300 & 3.80\% & 4.71\% & 4.97\% & 4.86\% & 3.92\% \\
            &   500 & 3.98\% & 4.69\% & 5.01\% & 4.92\% & 4.04\% \\
            &  2000 & 4.41\% & 4.85\% & 4.82\% & 4.95\% & 4.47\% \\
            \bottomrule 
    \end{tabular} 
}
\end{table}

\subsection{Synthetic data under alternative hypotheses}\label{sec:sim:H1}

This section provides an in-depth analysis of our testing procedure's power and evaluates the efficacy of the algorithm used for locating the first change point, employing synthetic data. We adopt the data structure described in Section~\ref{sec:simulation:mdl}, with the signal defined as per~\eqref{eqn:signal_flux} (i.e., \(H_1\)).

Our experimental setup is as follows:
we vary the sample size, choosing \(n\) from the values $50, 100, 300, 500, 2000$. We select the dependence parameter \(\theta\) from the values $-0.4$, $-0.2$, $0$ (representing independence), $0.2$, and $0.4$. The gap parameter \(s\) ranges from \(0\) to \(0.045\), increasing in steps of \(0.0006\). It is important to note that the standard deviation of the innovation in the dependent process is fixed at \( 0.5\), and we keep \(\mu_1 = 0\). We standardize the ratios \(\tau / n = 0.4\), \(\tau' / n = 0.6\), \(\tau'' / n = 0.8\), maintaining \(\mu_1 = 0\). For our testing methods, we consistently set the significance level at \(\alpha = 0.05\). Once the parameters for an experiment are established, we generate the trend \((\mu_i)\) using the aforementioned methodology. Subsequently, the additive noise process is simulated repeatedly, and this data is input into our testing and locating algorithms. For testing we use the quantile obtained from the asymptotic approximation and the true long-run variance $\sigma^2_{\infty}$ instead of $\hat\sigma^2_{\infty}$ in~\eqref{eq:def_T}; cf.\ Table~\ref{tab:sigma}. For our locating algorithm we first apply the test and continue only if it rejects. We use the long-run variance estimator defined in Section~\ref{sec:est_lrv}; i.\,e., $\hat\sigma_{\infty}^2 := \hat\sigma^2$; cf. \eqref{eq:lrv_est_mu0hat} and~\eqref{def:Dj}. The results are derived from 100,000 independent simulations. 

It is crucial to observe that testing for a change point remains challenging, even in scenarios with the largest gap parameter $s = 0.045$. This difficulty arises because, as indicated in Table~\ref{tab:sigma}, the gap $s = 0.045$ is considerably smaller than the noise levels, complicating the detection of the change point's presence significantly.

\begin{figure}[t]
	\centering
	\includegraphics[width=\textwidth]{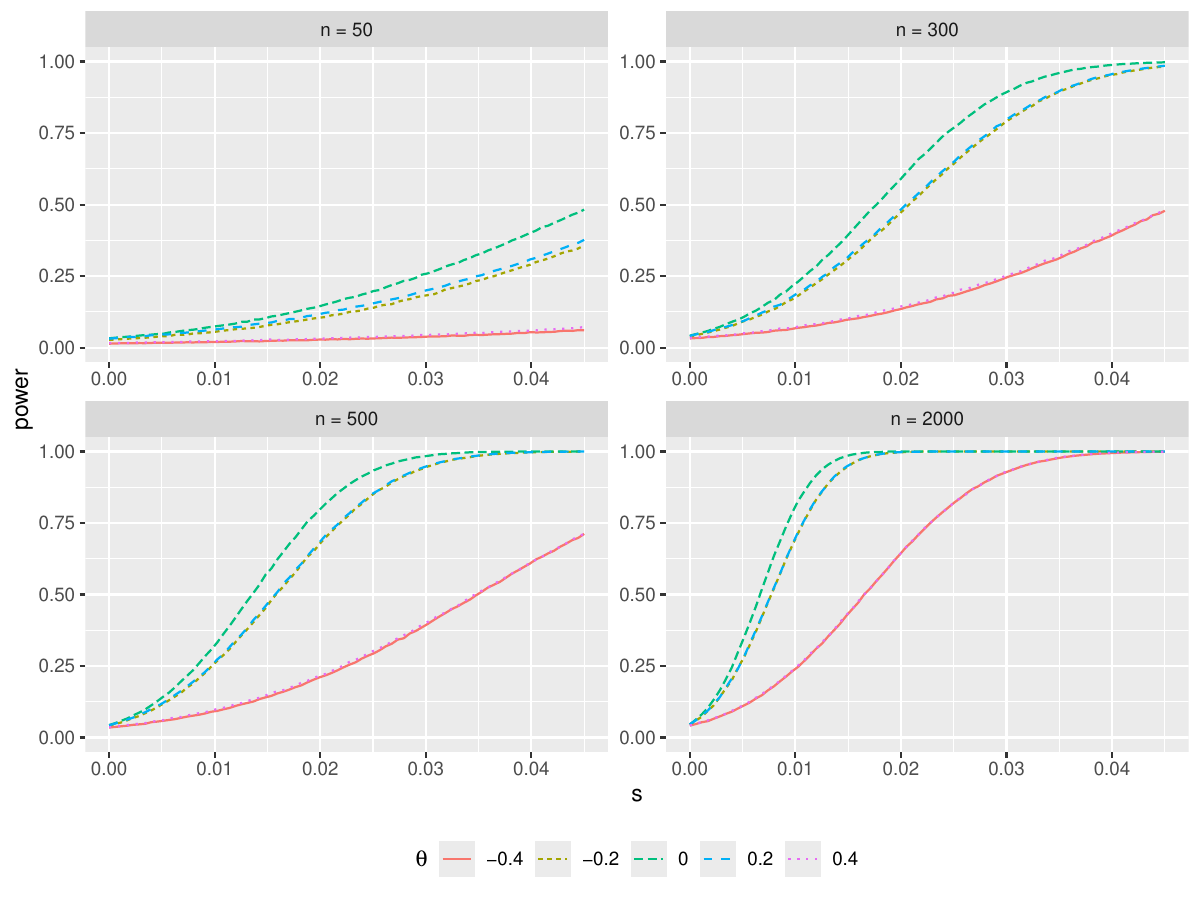}
	\caption{Rejection ratios for the testing procedure under the alternative hypothesis: \(\mu_i = \mu_1\) for \(i = 1,2,\ldots,\tau-1\); \(\mu_i > \mu_1 + s\) for \(i = \tau, \tau + 1, \tau + 2, \ldots,n\). The noise process is shaped by the dependence parameter \(\theta\). We adjust the gap parameter \(s\) over the set \(\{0, 0.0006, 0.0012, \ldots, 0.045\}\), \(n\) over \(\{50, 300, 500, 2000\}\), and \(\theta\) over \(\{-0.4, -0.2, 0, 0.2, 0.4\}\). Each data point represents 100,000 replications.}
	\label{Fig:ratio_alt}
\end{figure}

Initial observations indicate variations in the rejection ratio, an estimate of the true power, in relation to the gap parameter \(s\). These variations are evident across different combinations of the dependence parameter \(\theta\) and sample size \(n\), as depicted in Figure~\ref{Fig:ratio_alt}. In each experiment, the rejection ratio progresses from the nominal level (\(\alpha = 0.05\)) to nearly 1 as \(s\) increases from 0 to 0.045. This trend suggests that as the task of detecting change points becomes less challenging, the power of our test approaches unity.

The graphic shows an increase in the rejection ratio with sample sizes expanding from 50 to 2000. This trend is in alignment with the theoretical insights presented in Theorem~\ref{thm:size_test}(ii). Additionally, it is noteworthy that despite a diminished test power under conditions of strong temporal dependence (with \(|\theta| = 0.4\)), the power can still approach unity given a sufficient sample size. This observation implies the efficacy of our testing procedure even under the influence of temporal dependent noise in the data. Results for the case when the long-run variance is estimated are delegated to Appendix~B (in the Supplementary Material).

\begin{figure}[b]
	\centering
    \includegraphics[width=\textwidth]{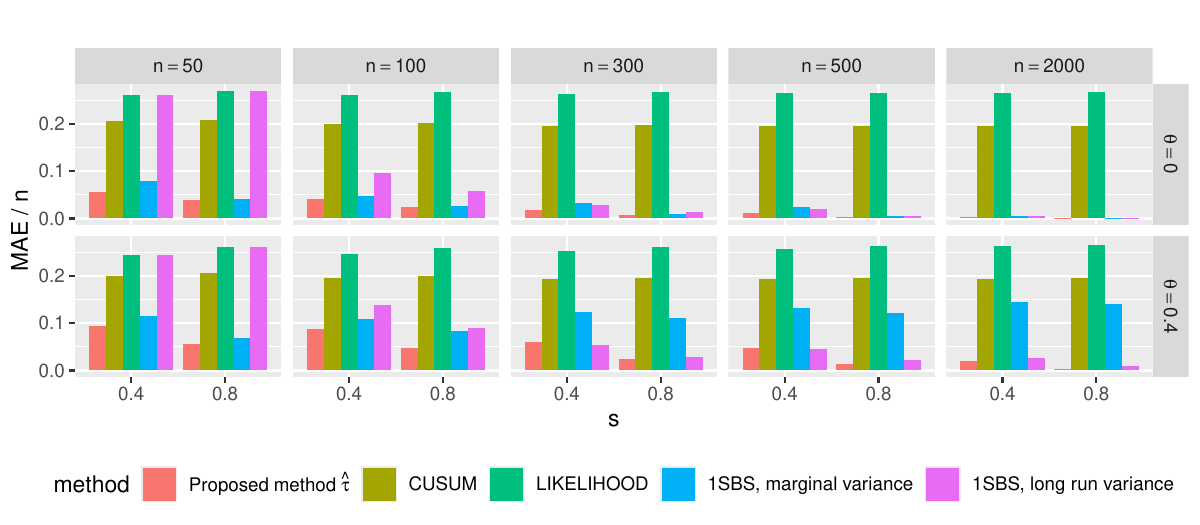} 
	\caption{\label{Fig:log_MAE}Expected absolute errors normalized by sample size (MAE/$n$) across four change point 
 detection methods. The red, olive, green, blue and purple bars show $\E |\hat{\tau} - \tau|/n$ of our proposed method $\hat\tau$, CUSUM, likelihood-based method (AMOC), the earliest change point from the standard binary segmentation (1SBS) method, and the earliest change point from the 1SBS method with marginal variance in the threshold replaced by long-run variance, respectively. The parameters varied in this study include the gap parameter $s$, the sample size $n$, and the parameter $\theta$ of the threshold autoregression noise process. Each bar in the graph represents the average result from $100,000$ replications.}
\end{figure}

Next, we showcase the absolute errors normalized by sample size \(\E |\hat{\tau} - \tau| / n\) of our two-step locating algorithm across experiments with diverse parameters in Figure~\ref{Fig:log_MAE}.

Without temporal dependence, error rates are smaller.  As the disparity between the signal and non-signal segments grows from 0.4 to 0.8, the error rate decreases. As the sample size~\(n\) expands from 50 to 2000 the MAE/$n$ progressively diminishes. These findings resonate with Theorem~\ref{thm:tauhat}, discussed in Section~\ref{sec:theory:loc_algo}.
For scenarios characterized by heightened dependence and minimal gap, error rates can be larger. Yet, in more favorable conditions, the error remains relatively stable or increases only marginally. This fact underscores the robustness of our methodology.

We expanded our analysis to compare the performance of our locating algorithm with four established change point estimation techniques: a CUSUM-type method, a likelihood-based method (AMOC), the earliest change point from the standard binary segmentation (1SBS) method, and a modified 1SBS method where the marginal variance in the threshold is replaced by the estimated long-run variance. More precisely, we obtain $\arg\min_{j = 2,3,\ldots,n+1} \sum_{i=1}^{j-1} (X_i - \bar{X}_n)$ and refer to it as CUSUM. This is related to our test statistic $\hat T$, defined in~\eqref{eq:def_T}. Secondly, we apply the functions \code{cpt.mean}, with \code{method = "AMOC"} (at most one change), and \code{cpts} from the \proglang{R} package \pkg{changepoint} \citep{changepointJSS,changepoint} and refer to the obtained value as AMOC.

Thirdly, we apply the functions \code{sbs} and \code{changepoints} from the \proglang{R} package \pkg{wbs} \citep{wbs} and refer to the minimum of the obtained values as 1SBS (first time of change obtained with standard binary segmentation). Fourthly, we replace the marginal variance used in the definition of the threshold of the standard binary segmentation method by the long-run variance, which we estimate by $\hat\sigma^2$, from Section~\ref{sec:est_lrv}, with $k = \lceil n^{1/3} \rceil$ and $J = 3$.

We add +1 to AMOC, 1SBS, and 1SBS based on long-run variance, to account for the fact that in our notation the change occurs from $\tau-1$ to $\tau$ while there it occurs from $\tau$ to $\tau+1$. Note that, while the first two methods estimate a single change point, the binary segmentation methods estimate multiple change points of which we select the earliest.

The outcome of this comparative study is detailed in Figure~\ref{Fig:log_MAE}. The results reveal that the errors associated with the four alternative methods are somewhat unstable and, depending on the scenario, can perform poorly.
The 1SBS method with the standard threshold performs roughly equally well under independence when the gap size is $s = 0.8$. When the gap size takes the smaller value $s=0.4$, where the locating problem is harder, the proposed method has slightly better performance. The 1SBS method with the threshold adjusted for long-run variance also performs well when there is serial dependence, but only for larger sample sizes.
We also note that keeping the gap parameter ($s$) and the dependence parameter ($\theta$) constant while increasing the sample size ($n$) from 50 to 2000 results in the mean absolute error (MAE/$n$) normalized by $n$ for our proposed method approaching zero. This observation confirms our theory and previous numerical analysis that our method's error remains relatively constant with larger sample sizes. In contrast, for the other four methods, we observe less stable behavior of the MAE/$n$, which either remains relatively unchanged as $n$ increases (CUSUM and AMOC), indicating that their errors grow with the sample size, or behave reasonably under independence but struggle in the presence of serial dependence (1SBS), or do not perform well for small sample sizes (1SBS with LRV).

Furthermore, when $\theta$ and $n$ are fixed and $s$ is varied, our method demonstrates a steady decline in error as $s$ increases. Such a consistent pattern of reducing error is also not observed in all of the other methods.

Overall, these results highlight the shortcomings of traditional methods in handling non-standard or complex data configurations, emphasizing the versatility of our proposed method. The bar plots in Figure~\ref{Fig:log_MAE} provide visual evidence of the consistently satisfactory performance of our method across various parameter settings, while the alternative methods exhibit unstable and sometimes poor performance, particularly in the presence of serial dependence.

\section{Baidu search index for COVID-19 related symptoms}

\label{sec:real_data}
\label{sec:da:baidu}

\begin{figure}[t]
    \centering
    \includegraphics[width=0.8\textwidth]{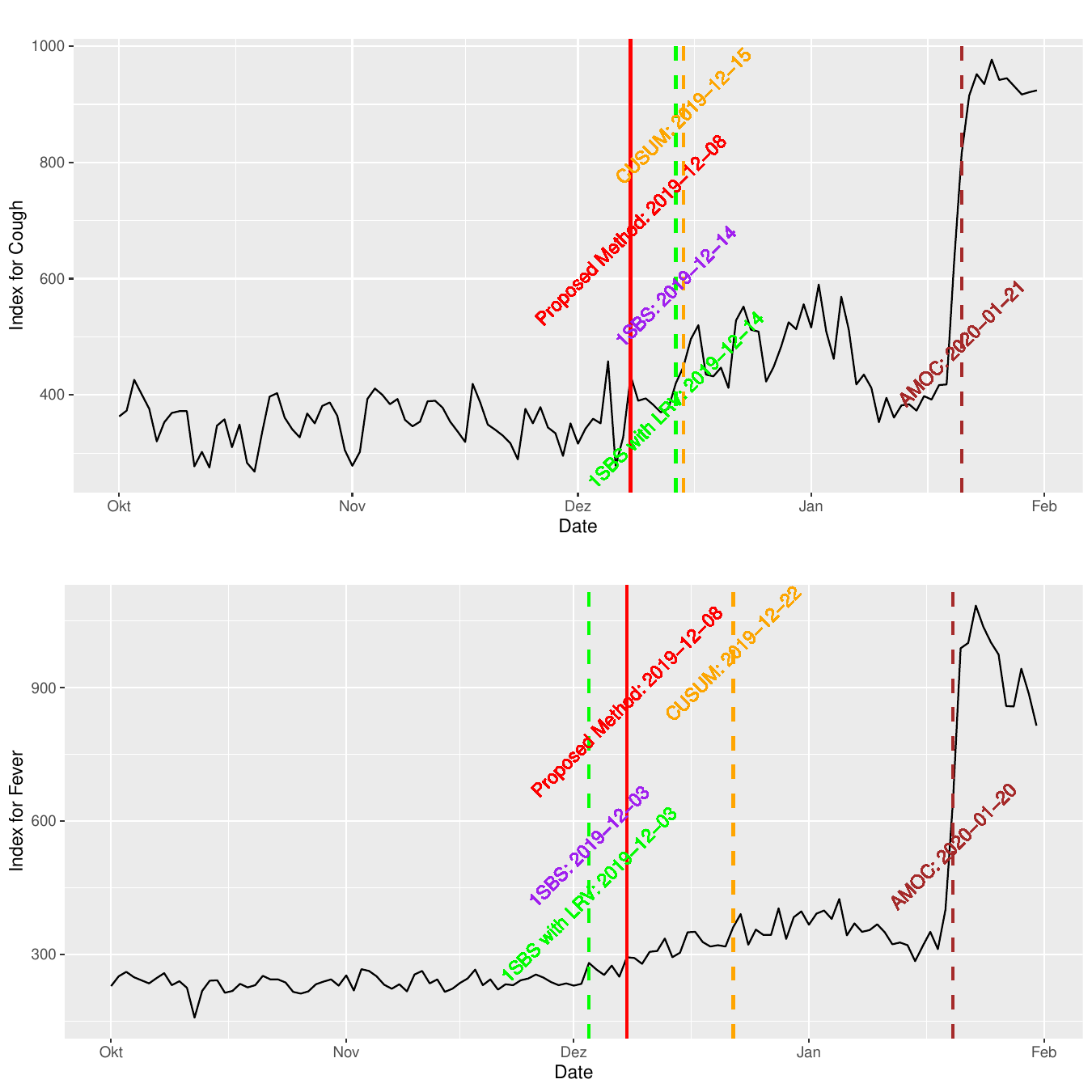} 
    \caption{BAIDU search index for ``fever'' and ``cough'' from 1 October 2019 to 31 January 2020.
    The red solid vertical line indicates the change point detected by our proposed method $\hat\tau$; the dashed lines in orange, brown, purple, and green represent the change points detected by CUSUM, likelihood-based (AMOC), earliest change detected by standard binary segmentation (1SBS), and binary segmentation with long-run variance modification (1SBS with LRV), respectively.}
    \label{Fig:covid_compare}   
\end{figure}

Numerous studies have endeavored to pinpoint the initial emergence of the SARS-CoV-2 virus among humans. The initial cases were likely linked to the Huanan Seafood Wholesale Market in late December 2019. However, this cluster is not believed to signify the pandemic's inception. To deduce the possible duration SARS-CoV-2 circulated in China before detection, we analyzed Baidu's search index (China's leading search engine) for COVID-19 symptom-related keywords between 1 October 2019 and 31 January 2020 in Hubei Province, China. We focused on the terms ``fever'' and ``cough'', aggregating searches from both desktop and mobile platforms. As depicted in Figure~\ref{Fig:covid_compare}, the counts exhibit regular fluctuations until the series' end. Given the rapid transmission capability of COVID-19, the constant mean assumption post-change point in conventional methods is inapplicable. Applying the test proposed in Section~\ref{sec:test} for the null hypothesis $H_0$ of constant mean, defined in~\eqref{eq:H0}, against the alternative hypothesis $H_1$ of a one-sided upwards change, defined in~\eqref{eq:H1}, yields test statistics $\hat T < -9$ and $\hat T < -22$, for Baidu search indices ``cough'' and ``fever'', respectively. The $p$-values implied by Theorem~\ref{thm:size_test}(i) are essentially zero such that we reject the null hypothesis in both cases.

We continue the analysis by employing the two-stage locating method (Section~\ref{sec:loc_algo}). For the keyword ``cough'' (comprising $n = 123$ data points), the initial stage estimates the equilibrium data state's mean, $\mu_1$, and the state gap parameter, $d$, guiding the subsequent stage. We defined $k = \lceil n^{1/3} \rceil = 5$ for the batched mean length and computed
\[
R_{j} = \frac{1}{k}\sum_{i = (j-1)k + 1}^{jk}X_j, \quad j =1, \ldots, m,
\]
as defined in~\eqref{eq:def_R}. We obtain $\hat L := \max\{i : R_i \leq R^{(3)}_m\} = 11$ and $\hat\ell := k \hat L = 55$, as defined in~\eqref{eq:def_L}.
We find a pre-change sample mean of $\hat{\mu}_0 = 352.84$ obtained from the initial $\hat\ell$ data. The test statistics $\hat{D}_j = \sqrt{k}(R_{j} - \hat{\mu}_0)/\hat{\sigma}_{\infty}$ using $\hat{\sigma}_{\infty} \approx  48.68$ which is the square root of the estimated long-term variance from the initial $\hat\ell$ observations; cf. \eqref{eq:lrv_est_mu0hat}. Then, we obtain the test decisions $\hat I_j$, defined in~\eqref{def:Dj}, as
\[
\hat{I}_j = 
\begin{cases} 
1 & \text{if } \hat{D}_j \ge z_{1-1/m} \\
0 & \text{otherwise,}
\end{cases}
\]
where $z_{1-1/m}$ is the $1-1/m$ quantile of the standard normal distribution.
We obtain
\[
\hat{\eta} := \argmin_t \sum_{j=1}^{m}\{I_j - 1_{[t+1, m]}(j)\}^2 = 15,
\]
as defined in~\eqref{def_eta}.
A graphical representation of the test decisions and smoothing can be seen in Figure~\ref{Fig:motivation_eta_hat}.
The first-stage estimates thus are
\[
\hat{\mu}_1 := \frac{1}{ k\hat{\eta}}\sum_{i=1}^{ k\hat{\eta}}X_i \approx 355.43, \quad
\hat{d} := \min_{\substack{i =  k (\hat\eta + 1) + 1,\\ \ldots, n-k+1}} \ \frac{1}{k} \sum_{j = i}^{i + k -1} (X_j - \hat{\mu}_1) \approx 19.24.
\]
Setting $\rho= 0.5$, our refined change point estimate in the second phase is:
\[
\hat\tau := \arg\min_{j = 2, \ldots, n} \sum_{t = 1}^{j-1} (X_t - \hat{\mu}_1 - \rho \hat{d} ) = 69,
\]
as defined in~\eqref{eq:def_tauhat}, which translates to 8 December 2019.

For comparative purposes, we also compute the CUSUM, AMOC and 1SBS estimates considered in Section~\ref{sec:sim:H1}. The CUSUM approach identified 15~December 2019 as the change point. The AMOC approach, pinpointed 21 January 2020 as the change point. In contrast, the Binary Segmentation method and its modified version, where the marginal variance in the threshold is replaced by the long-run variance, focusing on the first change point, suggests 14~December 2019 as the initial outbreak date, which precedes the dates indicated by the other two methods but is still later than our findings.

Interestingly and surprisingly, but reasonably, our analysis of the Baidu ``fever'' search index corroborates our findings by \emph{also} indicating 8 December 2019 as the change point, consistent with the ``cough'' dataset results. Conversely, the CUSUM, AMOC, and 1SBS methods suggest change points on 22 December 2019, 20 January 2020, and 3 December 2019, respectively. A graphical representation of the results can be seen in Figure~\ref{Fig:covid_compare}.

Reports such as \cite{worobey2021dissecting} mention an early COVID-19 case, a 41-year-old male, showing symptoms on 16 December 2019, suggesting community transmission. Another case, a female seafood vendor, exhibited symptoms on 10 December 2019 and was aware of potential COVID-19 cases near Huanan Market from 11 December 2019. Other studies and organizations like the CDC mention early December 2019 as significant. Given these findings, our change point detection appears plausible.

It is noteworthy that the Chinese government officially announced the outbreak on 20 January 2020, a discernible tipping point. This is not our primary focus, as our aim is to identify the initial outbreak, which undoubtedly predates 1 January 2020. Classical methods seem ill-equipped to discern this early change point, potentially overshadowed by subsequent tipping points. This is understandable, as such methods rely heavily on the sample mean, which can be skewed by later data points, leading to inaccurate estimations.

\begin{acks}[Acknowledgments]
 The authors would like to thank the Associate Editor and the three reviewers for their constructive comments that helped to improve the paper. This research is partially supported by NSF DMS-2311249 and NSF DMS-2027723. 
\end{acks}

\begin{supplement}

\stitle{Appendix}
\sdescription{Contains the proofs and additional simulation results.} 
\end{supplement}
\begin{supplement}
\stitle{Replication Package}
\sdescription{\proglang{R} code implementing the proposed method and scripts to replicate the simulation and empirical results in the paper are available on \url{https://github.com/tobiaskley/cp_analysis_w_irreg_signals_replication_package}.}
\end{supplement}

\bibliography{references.bib}

\end{document}


\begin{frontmatter}
\title{Supplementary Material to ``Change point analysis with irregular signals''}
\runtitle{Supplement to ``Change point analysis with irregular signals''}

\begin{aug}
\author[A]{\fnms{Tobias}~\snm{Kley}\ead[label=e1]{tobias.kley@uni-goettingen.de}},
\author[B]{\fnms{Yuhan Philip}~\snm{Liu}\ead[label=e2]{yuhanphilipliu@uchicago.edu}}
\author[C]{\fnms{Hongyuan}~\snm{Cao}\ead[label=e3]{hcao@fsu.edu}}
\and
\author[B]{\fnms{Wei Biao}~\snm{Wu}\ead[label=e4]{wbwu@uchicago.edu}}
\address[A]{Institute for Mathematical Stochastics, Georg-August-Universit\"at G\"ottingen, Germany\printead[presep={,\ }]{e1}}

\address[B]{Department of Statistics, University of Chicago, Chicago, IL, 60637\printead[presep={,\ }]{e2,e4}}
\address[C]{Department of Statistics, Florida State University, Tallahassee, FL, 32306\printead[presep={,\ }]{e3}}
\end{aug}

\begin{abstract}
This supplementary material contains the proofs and additional simulation results.
\end{abstract}

\begin{keyword}[class=MSC]
	\kwd[Primary ]{62M10}
	\kwd{62G20}
\end{keyword}

\begin{keyword}
\kwd{Change point analysis}
\kwd{COVID-19}
\kwd{invariance principle}
\kwd{irregular signals}
\kwd{long-run variance estimate}
\kwd{smoothing}
\kwd{weak convergence}
\end{keyword}

\end{frontmatter}

\appendix

\section{Proofs}
\label{sec:proofs}

We begin by introducing some notation and technical results used in the proofs.
During the proofs we will refer to the following conditions, which are satisfied under the conditions of the theorems: let the quantities $n$, $d$, $\tau$, $m$, $\eta$ (cf. comment before Theorem~3.2), and $\theta$, as in Condition~3.1, satisfy, as $n \to \infty$, that
\begin{itemize}
 \item[(C0)] $\eta \to \infty$,
 \item[(C1)] $n^{1/\theta} \sqrt{\log(\eta) / k} = \mathcal{O}(1)$,
 \item[(C2)] $( k d + \sqrt{k \log(\eta)} ) / n^{1/\theta} \rightarrow \infty$,
 \item[(C3)] $\log(m) \ll k d^2$,
 \item[(C4)] $m^{1/\theta} / k \ll k^{-1/2}$, as $n \to \infty$.
 \item[(C5)] $\hat\sigma^2_{\infty} / \sigma_{\infty}^2 \to 1$ in probability.
 \end{itemize}

By $\tau \leq (\eta + 1) k$, we have that $k \ll \tau$ implies (C0).
Further, we have that $k \gg n^{2/\theta} \log(n)$ and $d \gg n^{-1/\theta}$ imply (C1)--(C4). In Theorem~3.3 we required $\hat\sigma^2_{\infty} = \sigma^2_{\infty} + o_{\IP}(1)$, which implies (C5).
If $\hat\sigma_{\infty}^2$ is chosen to be our long-run variance estimate $\hat\sigma^2$, defined in~(8), then, under the the conditions of Theorem~3.2 (C5) is satisfied.

Throughout Sections~\ref{sec:proofs:thm:sigmahat}--\ref{sec:proofs:auxresults} we will apply Theorem~2 from \cite{WuWu2016} that provides a Nagaev-type inequalty.
The condition required to apply Theorem~2 from \cite{WuWu2016} is that the dependence adjusted norm, which is defined as
\begin{equation}
\label{eq:def_Xi}
\Xi_{\theta, \alpha} := \sup_{i \geq 0} (i+1)^{\alpha} \Theta_{i, \theta},
\end{equation}
is finite. In the definition of $\Xi_{\theta, \alpha}$, the cumulative dependence measure $\Theta_{i, \theta}$ is the same as the one we defined in~(15).

More precisely, to apply Theorem~2 in \cite{WuWu2016}, we require
\begin{condition}
	\label{cond:41}
	$(Z_i)_{i \in \Z}$ satisfies that the $\theta$-th moment $H_{\theta} :=  (\E |Z_i|^{\theta})^{1/\theta} < \infty$, where $\theta > 2$. Assume there exists $\alpha \geq 0$ such that $\Xi_{\theta, \alpha}  < \infty$.
\end{condition}
Clearly, we have that if Condition~3.1 holds, then Condition~\ref{cond:41} holds with $\alpha = 1$.

\subsection{Proof for Theorem~3.1}
\label{sec:proof:thms_test}

First, letting
\[T_n := \min_{j = 1,2,\ldots,n} \sum_{i=1}^j (X_i - \bar{X}_n) / (\sqrt{n} \sigma_{\infty})\]
and noting that by $\hat\sigma^2 = \sigma_{\infty}^2 + o_{\IP}(1)$ and Slutsky's Lemma, it suffices to prove the result with $\hat T$ replaced by $T_n$. Now, note that
\begin{equation}\label{J11804}
n^{-1/2} \sum_{i=1}^j (X_i - \bar X_n)
= n^{-1/2} M_n(j) + n^{-1/2} \Big( S_j - \frac{j}{n} S_n \Big), \quad S_n = \sum_{i=1}^n Z_i,
\end{equation}
where $M_n(j) := \sum_{i=1}^j (\mu_i - \bar \mu_n)$, $\bar\mu_n := n^{-1} \sum_{i=1}^n \mu_i$. By Theorem 3 in \cite{Wu2005}, under condition (16), we have the weak convergence
\begin{equation*}
n^{-1/2}  \{S_{\lfloor n u \rfloor}, 0 \le u \le 1 \}   
\Rightarrow \{ \sigma_\infty \mathbb{B}(u), 0 \le u \le 1 \}.    
\end{equation*}
Then $n^{-1/2}  \{ S_{\lfloor n u \rfloor} - n^{-1} \lfloor n u \rfloor S_n , 0 \le u \le 1 \} \Rightarrow \{ \sigma_\infty \mathbb{B}_1(u), 0 \le u \le 1 \}$, and case (i) follows from the continuous mapping theorem and (18), as under $H_0$, we have $M_n(j) = 0$.

For the proof of (ii), under $H_1$, for $j, \tau \in \{1, \ldots, n\}$, we have
\begin{equation}
\label{eqn:rep_Mn}
M_n(j) := \sum_{i=1}^j (\mu_i - \bar \mu_n) = \sum_{i=1}^j (\mu_i - \mu_1) - \frac{j}{n} \sum_{i = \tau}^{n} (\mu_i - \mu_1).
\end{equation}
The representation in~\eqref{eqn:rep_Mn} has interesting consequences. (a) Noting that $(\mu_i - \mu_1) = 0$, for $i=1, \ldots, \tau-1$, we have that the first term in the representation is non-negative and vanishes for $j < \tau$ and the second term is decreasing as $j$ increases, implying $\arg\min_j M_n(j) \geq \tau - 1$. (b) We have $\min_{j = 1,\ldots,n} M_n(j) \leq M_n(\tau-1) \leq d (\frac{1-\tau}{n})(n-\tau+1)$. Thus, \[n^{-1/2} \min_{j = 1,\ldots,n} M_n(j) \rightarrow -\infty, \quad \text{ as $n^{-3/2} \tau (n-\tau) d \rightarrow \infty$}.\]
Thus (ii) follows from (\ref{J11804}).
\hfill\qed

\noindent
\subsection{Proof of Theorem~3.2}
\label{sec:proofs:thm:sigmahat}

Lemma~\ref{lem:hatL}, below, is about the index $\hat L$ and asserts that, with probability tending to 1, $\hat L$ diverges and will not take values larger than $\eta - o(\eta)$.
\begin{lemma}
	\label{lem:hatL}
    Grant Conditions~3.1, (C1), (C2), and (C3).
    Let $\hat L = \max\{i : R_i \leq R^{(J)}_m\}$, for fixed $J \in \N$, be the generalized version of $\hat L$.
    Then, for any sequence $0 \leq t_n = o(\eta)$,
\begin{equation}
\label{eq:asymptotic_L_n}
\IP(t_n \leq \hat{L} \leq \eta - t_n) \to 1, \quad \text{as $n \to \infty$}.
\end{equation}
 \end{lemma}

The proof is deferred to Section~\ref{sec:proofs:hatL}.

\begin{lemma}
\label{lem:smSpecDens}
Let $\Theta_{n, 2}$ be defined as in~(15). Then
    \begin{equation}
    \label{eq:smSpecDens}
        \sum_{k=-\infty}^{\infty} |\gamma(k)| \leq \Theta_{0, 2}^2, \quad
        \sum_{k=-\infty}^{\infty} |k| |\gamma(k)| \leq 2 \Theta_{0, 2} \sum_{i = 0}^{\infty} \Theta_{i, 2}.
    \end{equation}
In particular, Condition~3.1 implies that $\sum_{k=-\infty}^{\infty} |k| |\gamma(k)| < \infty$.
\end{lemma}
The proof is deferred to Section~\ref{sec:proofs:smSpecDens}.

\clearpage
Lemma~\ref{lem:rate_|hat_mu1-mu1|}, below, is about the consistency and rate of the estimate $\hat\mu_0$ that is also used in the definition of the test statistic $\hat{D}_j$.
\begin{lemma}
	\label{lem:rate_|hat_mu1-mu1|}
	Grant Conditions~3.1, (C1), (C2), and (C3). Then
	\begin{equation}
	\label{eq:rate_|hat_mu1-mu1|}
	|\hat{\mu}_0 - \mu_1| = O_\IP\Big(\frac{1}{\sqrt{\eta k}} \Big), \quad \text{as $n \to \infty$}.
	\end{equation}
\end{lemma}
The proof is deferred to Section~\ref{sec:proofs:rate_|hat_mu1-mu1|}.

Now we prove Theorem~3.2. The following decomposition holds:
\begin{align}
& \hat{\sigma}^2 - \sigma_{\infty}^2 \nonumber \\
& = \hat{\sigma}^2 - \frac{k}{k \hat{L} - k + 1} \sum_{s=k}^{k \hat{L}} \Big( R_{s/k} - \mu_1 \Big)^2 \label{eqn:repr1_sigmahat} \\
    & \quad + \frac{k}{k \hat{L} - k + 1} \sum_{s=k}^{k \hat{L}} \Big( R_{s/k} - \mu_1 \Big)^2
 - \frac{k}{k \hat{L} - k + 1} \sum_{s=k}^{k \hat{L}} \Big( R_{s/k} - \IE R_{s/k} \Big)^2 \label{eqn:repr2_sigmahat} \\
 & \quad + \frac{k}{k \hat{L} - k + 1} \sum_{s=k}^{k \hat{L}} \Big( R_{s/k} - \IE R_{s/k} \Big)^2 - k  \IE \Big( R_1 - \IE R_1 \Big)^2 \label{eqn:repr3_sigmahat} \\
 & \quad + k \IE \Big( R_1 - \IE R_1 \Big)^2 - \sigma_{\infty}^2 \label{eqn:repr4_sigmahat}
\end{align}
We will show that
\[\eqref{eqn:repr1_sigmahat} = \mathcal{O}_{\IP}(1/\eta), \,\,
\eqref{eqn:repr2_sigmahat} = \mathcal{O}_{\IP}\big( (\eta k)^{-1/2} \big), \,\,
\eqref{eqn:repr3_sigmahat} = \mathcal{O}_{\IP}( \eta^{2/\min(4, \theta)-1} \big),
\,\, \text{and} \,\, 
\eqref{eqn:repr4_sigmahat} = \mathcal{O}(1/k).
\]

To show $\eqref{eqn:repr1_sigmahat} = \mathcal{O}_{\IP}(1/\eta)$, it suffices to show that $\eqref{eqn:repr1_sigmahat} = \mathcal{O}_{\IP}(a_n/\eta)$ for any $a_n \to \infty$. Let $t_n := \eta/a_n$, then we have $t_n = o(\eta)$. Now, for~\eqref{eqn:repr1_sigmahat} note that
\begin{align}
& \Big| \hat{\sigma}^2 - \frac{k}{k \hat{L} - k + 1} \sum_{s=k}^{k \hat{L}} \Big( R_{s/k} - \mu_1 \Big)^2 \Big| \nonumber \\
& =  \Big| \frac{k}{k \hat{L} - k + 1} \sum_{s=k}^{k \hat{L}} \Big( R_{s/k} - \hat\mu_0 \Big)^2 - \frac{k}{k \hat{L} - k + 1} \sum_{s=k}^{k \hat{L}} \Big( R_{s/k} - \mu_1 \Big)^2 \Big| \nonumber  \\
& = \Big| \frac{2 k}{k \hat{L} - k + 1} \sum_{s=k}^{k \hat{L}} (R_{s/k} - \IE R_{s/k}) ( \mu_1 - \hat\mu_0) \nonumber \\
& \qquad\qquad + \frac{k}{k \hat{L} - k + 1} \sum_{s=k}^{k \hat{L}} \Big( (2 \IE R_{s/k}) ( \mu_1 - \hat\mu_0) + \hat\mu_0^2 - \mu_1^2 \Big) \Big| \nonumber  \\
& = \Big| \frac{2 k}{k \hat{L} - k + 1} \sum_{s=k}^{k \hat{L}} (R_{s/k} - \IE R_{s/k}) ( \mu_1 - \hat\mu_0) \nonumber \\
& \qquad\qquad + \frac{k}{k \hat{L} - k + 1} \sum_{s=k}^{k \hat{L}} \Big( ( \hat\mu_0 - \mu_1)^2 + 2 (\mu_1 - \IE R_{s/k}) ( \hat\mu_0 - \mu_1 ) \Big) \Big| \nonumber \\
& \leq 2k | \mu_1 - \hat\mu_0 | \, \Bigg( \Big| \frac{1}{k \hat{L} - k + 1}\sum_{s=k}^{k \hat{L}} (R_{s/k} - \IE R_{s/k}) \Big|
+  \frac{1}{2} | \hat\mu_0 - \mu_1 | \nonumber  \\
& \hspace{5cm} + \frac{1}{k \hat{L} - k + 1} \sum_{s=k}^{k \hat{L}} |\mu_1 - \IE R_{s/k}| \Bigg) \nonumber \\
& = \mathcal{O}_{\IP} \Bigg( k \frac{1}{(\eta k)^{1/2}} \Big( \frac{ \eta/t_n}{(\eta k)^{1/2}} + \frac{1}{(\eta k)^{1/2}} + \frac{1}{(\eta k)^{1/2}} \Big)  \Bigg) = \mathcal{O}_{\IP} \Big( \frac{a_n}{\eta} \Big), \label{eqn:repr1_sigmahat:rate}
\end{align}
where we have used the fact that $| \hat\mu_0 - \mu_1 | = \mathcal{O}_{\IP}\big( (\eta k)^{-1/2} \big)$, by Lemma~\ref{lem:rate_|hat_mu1-mu1|} and, which we will prove below, that
\begin{equation}\label{thm:sigmahat:A}
    \frac{1}{k \hat{L}} \sum_{s=k}^{k \hat{L}} (R_{s/k} - \IE R_{s/k}) = \mathcal{O}_{\IP}\Big( \frac{(\eta k)^{1/2}}{k t_n} \Big),
\end{equation}
for all sequences $t_n = o(\eta)$, and
\begin{equation}\label{thm:sigmahat:B}
    \frac{1}{k \hat{L}} \sum_{s=k}^{k \hat{L}} |\mu_1 - \IE R_{s/k}| = \mathcal{O}_{\IP}\big( (\eta k)^{-1/2} \big),
\end{equation}
which together with $(k\hat L)/(k\hat L - k + 1) = 1 + o_P(1)$, as $n \to \infty$, which follows from Lemma~\ref{lem:hatL}, yields the rate in~\eqref{eqn:repr1_sigmahat:rate}.
For the proof of~\eqref{thm:sigmahat:A} first note that
\begin{equation*}
    \begin{split}
        & \max_{t_n \leq L \leq \eta} \Big| \frac{1}{k} \sum_{s=k}^{k L} \sum_{i=s-k+1}^{s} Z_i \Big|
         = \max_{t_n \leq L \leq \eta} \Big| \frac{1}{k} \sum_{u=1}^{k} \sum_{i=u}^{k(L-1)+u} Z_i \Big| \\
        & \leq \max_{t_n \leq L \leq \eta} \max_{u = 1,\ldots, k} \Big| \sum_{i=u}^{k(L-1)+u} Z_i \Big|
        = \max_{t_n \leq L \leq \eta} \max_{u = 1,\ldots, k} \Big| \sum_{i=1}^{k(L-1)+u} Z_i  - \sum_{i=1}^{u-1} Z_i \Big| \\
        & \leq 2 \max_{n = 1, \ldots, k \eta} \Big| \sum_{i=1}^{n} Z_i \Big|.
    \end{split}
\end{equation*}
Thus, letting $r_n := 2 (\eta k)^{1/2}/ (k t_n)$, we have
\begin{equation*}
\begin{split}
& \IP\Big( \Big| \frac{1}{k \hat{L}} \sum_{s=k}^{k \hat{L}} (R_{s/k} - \IE R_{s/k}) \Big| > M r_n \Big)
\leq \IP\Big( \max_{t_n \leq L \leq \eta} \Big|  \frac{1}{k L} \sum_{s=k}^{k L} \frac{1}{k} \sum_{i=s-k+1}^{s} Z_i \Big| > M r_n \Big) + o(1) \\
& \leq \IP\Big( \max_{1 \leq n \leq \eta k} \Big| \sum_{j=1}^{n} Z_j \Big| > M k t_n r_n / 2 \Big) + o(1) \\
& \leq C_{1} \frac{ \eta k \Xi_{\theta, 1}^{\theta} }{( M k t_n r_n / 2 )^{\theta}}+ C_2 \exp \left( - \frac{C_3}{\Xi_{2, 1}^{2}} \frac{ ( M k t_n r_n / 2  )^2 }{ \eta k }  \right) + o(1) \\
& = C_{1} \Xi_{\theta, 1}^{\theta} \frac{ 1 }{ M^{\theta} (\eta k)^{(\theta - 2)/2}  } + C_2 \exp \left( - \frac{C_3}{\Xi_{2, 1}^{2}} M^2  \right) + o(1),
\end{split}
\end{equation*}
which will be arbitrarily small for $M$ large enough. For the $o(1)$ after the first inequality in the above, we have used Lemma~\ref{lem:hatL}. For the third inequality, note that Condition~\ref{cond:41} holds. Therefore, we can apply Theorem~2 in \cite{WuWu2016}, and hence, there exist positive constants $C_1, C_2$ and $C_3$, such that the third inequality holds.

For the proof of~\eqref{thm:sigmahat:B}, we will prove the following, more general, statement:
let $W_1, \ldots, W_n$ be any random variables that are such that $W_k = \ldots = W_{k \eta} = 0$ a.\,s., and $w_n > 0$ an arbitrary sequence signifying a rate. Then,
\begin{equation}
\label{eqn:sumToHatL}
\sum_{i=k}^{k \hat L} W_i = o_{\IP}(w_n).
\end{equation}
To see this, note that for $\varepsilon > 0$
\begin{equation*}
\begin{split}
\IP\Big( \Big| \sum_{i=k}^{k \hat{L}} W_i \Big|  > \varepsilon w_n \Big)
& \leq \IP\Big( \Big\{ \Big| \sum_{i=k}^{k \hat{L}} W_i \Big| > \varepsilon w_n \Big\}  \cap \Big\{  \hat L \leq \eta \Big\} \Big) + \IP( \hat L > \eta ) \\
& \leq  \IP\Big( \sup_{L = 1, \ldots, \eta} \Big| \sum_{i=k}^{k L} W_i \Big|  > \varepsilon w_n \Big) + o(1) \to 0,
\end{split}
\end{equation*}
due to $\sup_{L=1, \ldots, \eta} \sum_{i=k}^{k L} W_i = 0$ a.\,s., and $o(1)$ follows from Lemma~\ref{lem:hatL}.
Thus, \eqref{thm:sigmahat:B} follows from~\eqref{eqn:sumToHatL} with $W_i = |\mu_1 - \IE R_{i/k}| / (k \hat{L})$ and $w_n = (\eta k)^{-1/2}$. We conclude $\eqref{eqn:repr1_sigmahat} = \mathcal{O}_{\IP}(1/\eta)$.

For~\eqref{eqn:repr2_sigmahat}, note that
\begin{equation*}
\begin{split}
 & \frac{k}{k \hat{L} - k + 1} \sum_{s=k}^{k \hat{L}} \Big( R_{s/k} - \mu_1 \Big)^2
 - \frac{k}{k \hat{L} - k + 1} \sum_{s=k}^{k \hat{L}} \Big( R_{s/k} - \IE R_{s/k} \Big)^2 \\
 & = \frac{k \hat L}{k \hat{L} - k + 1} \sum_{s=k}^{k \hat{L}} \frac{1}{\hat{L}} \Big( 2 R_{s/k} ( \IE R_{s/k} - \mu_1 ) + \mu_1^2 - (\IE R_{s/k})^2 \Big) = \mathcal{O}_{\IP}\big( (\eta k)^{-1/2} \big),
\end{split}
\end{equation*}
with the rate following from~\eqref{eqn:sumToHatL}.

For~\eqref{eqn:repr4_sigmahat}, note that
\begin{equation*}
\begin{split}
\sigma_{\infty}^2 - k \IE \Big( R_1 - \IE R_1 \Big)^2 
& = \sum_{u=-\infty}^{\infty} \gamma(u) - \sum_{|u| < k} (1 - |u|/k) \gamma(u) \\
& = \sum_{|u| \geq k } \gamma(u) + k^{-1} \sum_{|u| < k} |u| \gamma(u) = \mathcal{O}(1/k),
\end{split}
\end{equation*}
where the rate of $\mathcal{O}(1/k)$ follows from $\sum_{u=-\infty}^{\infty} |u \gamma(u)| < \infty$.

Now, to show $\eqref{eqn:repr3_sigmahat} = \mathcal{O}_{\IP}\big( \eta^{2/\min(4, \theta)-1} \big)$, by Lemma \ref{lem:hatL}, it suffices to prove that 
\begin{equation}\label{eq:temp48o}
    {1\over k \eta} 
    \left \| \max_{i \le \eta} \left | \sum_{s=k}^{k i} Y_s \right| \right \|_\phi = O\big( \eta^{2/\min(4, \theta)-1} \big), 
\end{equation}
where $\phi = \theta/ 2$ and 
\begin{equation}
\label{eqn:def_Yt}
Y_t := \tilde Y_t - \IE \tilde Y_t, \quad \tilde Y_t := \frac{1}{k} ( Z_{t-k+1} + \cdots + Z_t)^2 
 = k ( R_{t/k} - \IE R_{t/k} )^2.
\end{equation}
By the stationarity of $Y_t$, to prove (\ref{eq:temp48o}), it suffices to show that
\begin{equation}\label{eq:temp48}
    {1\over \eta} 
    \left \| \max_{h \le \eta} \left | \sum_{j=1}^{h} Y_{j k} \right| \right \|_\phi 
    = O\big( \eta^{2/\min(4, \theta)-1} \big). 
\end{equation}
To bound $\| Y_k \|_\phi$, let $S_k = Z_1 + \ldots + Z_k$. By Theorem~3 in \cite{wu2011asymptotic}, there exists  a constant $c_\theta > 0$ that depends only on $\theta$, such that $\| k^{-1/2} S_k \|_\theta \leq c_\theta \Theta_{0,\theta}$. Then 
\begin{equation}\label{eqn:bnd:norm:Yk}
\| Y_k \|_\phi \le 2 \| \tilde Y_k \|_\phi \le 2 c_{\theta}^2 \Theta_{0, \theta}^2.
\end{equation}
Further, we define, along the lines of~(14),
\begin{equation*}
    \begin{split}
        Z_{i,\{h\}} & := G(\ldots, \varepsilon_{h-1}, \varepsilon_{h}',\varepsilon_{h+1}, \ldots, \varepsilon_{i} ), \\
        S_{k,\{h\}} & := Z_{1,\{h\}} + \ldots + Z_{k,\{h\}}, \\
        Y_{i,\{h\}} & := S_{k,\{h\}}^2 / k.
    \end{split}
\end{equation*}
Then, we have
\begin{eqnarray}\label{eqn:bnd:PtYk}
\|  \mathcal{P}_t Y_{k}  \|_\phi &\leq& \| Y_{k} - Y_{k,\{t\}} \|_\phi
      = \frac{1}{k} \|  S_{k}^2 - S_{k,\{t\}}^2 \|_\phi \cr
        &\leq& \frac{1}{k} \|  S_{k} - S_{k,\{t\}} \|_{\theta}
        \|  S_{k} + S_{k,\{t\}}  \|_{ \theta} \cr
        & \leq& \frac{1}{k} \sum_{j=1}^k\| Z_{j} - Z_{j,\{t\}}\|_{\theta} (\|  S_{k}\|_{\theta} + \|  S_{k,\{t\}} \|_{ \theta} ) \cr
        & \leq& \frac{2}{k^{1/2}} \sum_{j=1}^k \delta_{ \theta}(j-t) \| k^{-1/2}  S_{k} \|_{ \theta}
        \leq \frac{ 2 c_{\theta} \Theta_{0, \theta}}{k^{1/2}} \sum_{j=1}^k \delta_{ \theta}(j-t), 
\end{eqnarray}
where the first inequality is due to Theorem~1 in \cite{Wu2005}, the second inequality is due to Cauchy-Schwarz inequality, the third inequality employs the definition of $S_k$ and $S_{k,\{t\}}$, as well as the triangle inequality, the fourth inequality employs the definition of $\delta_{\theta}$ and the fact that $S_k$ and $S_{k,\{t\}}$ are identically distributed. Finally, for the fifth inequality we have used Theorem~3 in \cite{wu2011asymptotic}, as before.

Let $\xi_t = (\varepsilon_t, \varepsilon_{t-1}, \ldots)$. Define the projection operator $\mathcal{P}_t \cdot = \IE( \cdot | \xi_t) - \IE( \cdot | \xi_{t-1})$. Then
\begin{equation}\label{temp43}
 Y_{\ell k} = Y_{\ell k} - \IE[ Y_{\ell k} | \xi_{ (\ell-1)k } ] + \IE[ Y_{\ell k} | \xi_{ (\ell-1)k } ]
  =  \{Y_{\ell k} - \IE[ Y_{\ell k} | \xi_{ (\ell-1)k } ] \} 
   + \sum_{i=k}^\infty \mathcal{P}_{\ell k -i} Y_{\ell k}.
\end{equation}
Let $\iota = \min(2, \phi)$. Since $Y_{\ell k} - \IE[ Y_{\ell k} | \xi_{ (\ell-1)k } ]$, $\ell \in \mathbb Z$, form stationary martingale differences, by Burkholder's inequality, there exists a constant $C_\phi$ only depending on $\phi$ such that
\begin{equation}\label{eq:temp491}
    \left \| \max_{h \le \eta} \left | \sum_{\ell =1}^{h} \{Y_{\ell k} - \IE[ Y_{\ell k} | \xi_{ (\ell-1)k } ] \} \right| \right \|_\phi^\iota
    \le C_\phi \eta \| Y_{k} - \IE[ Y_{ k} | \xi_{ 0 } ] \| _\phi^\iota
    \le C_\phi \eta (2 c_\theta \Theta_{0,\theta})^\theta. 
\end{equation}
Since $\mathcal{P}_{\ell k -i} Y_{\ell k}$, $\ell \in \mathbb Z$, are also stationary martingale differences, by Burkholder's inequality, 
\begin{equation}\label{eq:temp492}
    \left \| \max_{h \le \eta} \left | \sum_{\ell =1}^{h} \sum_{i=k}^\infty \mathcal{P}_{\ell k -i} Y_{\ell k} \right| \right \|_\phi
    \le  \sum_{i=k}^\infty \left \| \max_{h \le \eta} \left | \sum_{\ell =1}^{h}  \mathcal{P}_{\ell k -i} Y_{\ell k} \right| \right \|_\phi \le \sum_{i=k}^\infty  (C_\phi \eta) ^{1/\iota} \| \mathcal{P}_{ k -i} Y_{k} \|_\phi
\end{equation}
In (\ref{eqn:bnd:PtYk}) let $t = k - i$. By Conditions~3.1,
\begin{equation}\label{eq:temp493}
 \sum_{i=k}^\infty   \| \mathcal{P}_{ k -i} Y_{k} \|_\phi \le  \sum_{i=k}^\infty  
 \frac{ 2 c_{\theta} \Theta_{0, \theta}}{k^{1/2}} \sum_{j=1}^k \delta_{ \theta}(j-k+i)
 = \frac{ 2 c_{\theta} \Theta_{0, \theta}}{k^{1/2}} \sum_{j=1}^k \Theta_{j, \theta} = {\cal O}(k^{-1/2}).
\end{equation}
Then (\ref{eq:temp48}) follows from (\ref{temp43}), (\ref{eq:temp491}), (\ref{eq:temp492}) and (\ref{eq:temp493}), which complete the proof.

\subsection{Proof of Theorem~3.3}
\label{sec:proofs:thm:tauhat}

Our proof of the theorem makes use of the statistical properties of the estimate $\hat\eta$, defined in~(10), $\hat\mu_1$, defined in~(11), and $\hat d$, defined in~(11).
We summarize the properties of these estimates in Theorem~\ref{thm:etahat} and Lemmas~\ref{lem:mu1hat} and~\ref{lem:dhat}, below.
\begin{thm}\label{thm:etahat}
	Assume Conditions~3.1, and (C0)--(C5).
	 Then,
	\[
	\IP(\hat\eta = \eta)  \to 1, \quad n\to \infty.
	\]
\end{thm}
\noindent
The proof of Theorem~\ref{thm:etahat} is deferred to Section~\ref{sec:proofs:thm:etahat}.
For the proof of Theorem~3.3 we further employ the following lemmas
\begin{lemma}
	\label{lem:mu1hat}
    Under the condition of Theorem~\ref{thm:etahat}, we have
    \begin{equation}
    \label{eqn:rate_mu1hat}
        | \hat\mu_1 - \mu_1 | = \mathcal{O}_{\IP}\big( (k \eta)^{-1/2} \big).
    \end{equation}
\end{lemma}
The proof of Lemma~\ref{lem:mu1hat} is deferred to Section~\ref{sec:proofs:mu1hat}.

\begin{lemma}
    \label{lem:dhat}
    Under the condition of Theorem~\ref{thm:etahat} and $n - \tau \geq 2k$, we have
    \begin{equation}
        \label{eqn:rate_dhat}
        | \hat d - d_* | = \mathcal{O}_{\IP} \left( \Big( \frac{\log( m-\eta+1 )}{k} \Big)^{1/2} \right).
    \end{equation}
\end{lemma}
The proof of Lemma~\ref{lem:dhat} is deferred to Section~\ref{sec:proofs:dhat}.

\bigskip

Now, to prove $|\hat{\tau}_n - \tau| = \mathcal{O}_{\IP}\big(d_*^{-\theta/(\theta-1)}\big)$, as $n \to \infty$, we need to show that for any $\varepsilon > 0$, there exists $\tilde M_{\varepsilon} \in \N$ and $N_{\varepsilon} \in \N$ such that
$$
\IP\left(\left|\hat{\tau}_n - \tau\right| \geq \tilde M_{\varepsilon} d_*^{-\theta/(\theta-1)} \right) < \varepsilon, \quad \forall n > N_{\varepsilon}.
$$

We now derive a bound for $\IP\left(\left|\hat{\tau}_n-\tau\right| \geq M_{\varepsilon}\right)$, where $M_{\varepsilon} := \lfloor \tilde M_{\varepsilon} d_*^{-\theta/(\theta-1)} \rfloor$. Note that from~(12) we have that
\begin{align*}
\Omega &= \left\{  \sum_{t = 1}^{\tau - 1} (X_t - \hat\mu_1 - \rho\hat d) \geq \sum_{t = 1}^{\hat{\tau}_n - 1} (X_t - \hat\mu_1 - \rho \hat d) \right\} \\
& = \left(\left\{\sum_{t=\hat{\tau}_n}^{\tau-1} (X_t - \hat\mu_1 - \rho\hat d) \geq 0\right\} \cap\left\{\hat{\tau}_n<\tau\right\}\right) \\ 
& \qquad \cup\left(\left\{\sum_{t=\tau}^{\hat{\tau}_n-1} (X_t - \hat\mu_1 - \rho\hat d) \leq  0\right\} \cap\left\{\hat{\tau}_n>\tau\right\}\right) \cup\left\{\hat{\tau}_n=\tau\right\}.
\end{align*}

Then for $n > N_{\varepsilon} \gg M_{\varepsilon} > 0,$ we have the bound as follows
\begin{align}
&\IP\left(\left|\hat{\tau}_n-\tau\right| \geq M_{\varepsilon}\right)=\IP\left(\bigcup_{\ell=M_{\varepsilon}}^{\tau - 2}\left\{\hat{\tau}_n=\tau-\ell\right\} \cap \Omega\right)+\IP\left(\bigcup_{\ell=M_{\varepsilon}}^{n-\tau}\left\{\hat{\tau}_n=\tau+\ell\right\} \cap \Omega\right) \nonumber  \\
&=\IP\left(\bigcup_{\ell=M_{\varepsilon}}^{\tau - 2}\left\{\sum_{t=\tau-\ell}^{\tau-1}\left( X_t - \hat\mu_1 - \rho\hat d \right) \geq 0\right\}\right) \nonumber \\
& \quad +\IP\left(\bigcup_{\ell=M_{\varepsilon}}^{n-\tau}\left\{\sum_{t=\tau}^{\tau+\ell-1}\left( X_t - \hat\mu_1 - \rho\hat d \right) \leq 0\right\}\right) \nonumber \\
& \leq \IP\left(\max _{\ell \geq M_{\varepsilon}} \sum_{t=\tau-\ell}^{\tau-1}\left(Z_t + \mu_t - \hat\mu_1 - \rho \hat d\right) \geq 0\right)  \label{eq:P1} \\
& \quad + \IP\left(\max _{\ell \geq M_{\varepsilon}}
\sum_{t=\tau}^{\tau+\ell-1}\left( \hat\mu_1 + \rho \hat d - Z_t -  \mu_t  \right) \geq 0\right). \label{eq:P2}
\end{align}

We treat~\eqref{eq:P1} and~\eqref{eq:P2} separately.

For~\eqref{eq:P1}, note that $\mu_t = \mu_1$ for $t < \tau$.
Then, for any $\varepsilon$ with $0 < \varepsilon  < \rho d_* / (1 + \rho)$, we argue as follows:
\begin{equation*}
    \begin{split}
        \eqref{eq:P1} & = \IP\left(\max _{\ell \geq M_{\varepsilon}} \sum_{t=\tau-\ell}^{\tau-1} Z_t + \ell \Big( \mu_1 - \hat\mu_1 - \rho \hat d \Big)  \geq 0\right) \\
        & \leq \IP\left(\max _{\ell \geq M_{\varepsilon}} \sum_{t=\tau-\ell}^{\tau-1} Z_t + \ell \Big( (1 + \rho) \varepsilon - \rho d_* \Big)  \geq 0\right) + \IP\Big( | \hat d - d_* | > \varepsilon \Big) + \IP\Big( | \hat\mu_1 - \mu_1 | > \varepsilon \Big).        
    \end{split}
\end{equation*}
Further, for the first term, we have
\begin{equation*}
    \begin{split}
& \IP\left(\max _{\ell \geq M_{\varepsilon}} \sum_{t=\tau-\ell}^{\tau-1} Z_t + \ell \Big( (1 + \rho) \varepsilon - \rho d_* \Big)  \geq 0\right) \\
& \leq \sum_{k=1}^{\infty} \IP\left(\max _{2^{k-1}M_{\varepsilon} \leq \ell \leq 2^k M_{\varepsilon}} \sum_{t=\tau-\ell}^{\tau-1} Z_t + \ell \Big( (1 + \rho) \varepsilon - \rho d_* \Big)  \geq 0\right) \\
& \leq \sum_{k=1}^{\infty} \IP\left(\max _{1 \leq \ell \leq 2^k M_{\varepsilon}} \sum_{t=\tau-\ell}^{\tau-1} Z_t \geq 2^{k-1}M_{\varepsilon} \Big( \rho d_* - (1 + \rho) \varepsilon \Big) \right) \\
& \leq \sum_{k=1}^{\infty}  C_1 \frac{2^k M_{\varepsilon} \Xi_{\theta, 1}^{\theta}}{\Big( 2^{k-1}M_{\varepsilon} r_1 \Big)^{\theta}} +  \sum_{k=1}^{\infty}  C_2 \exp\Big(- \frac{C_3 \Big( 2^{k-1} M_{\varepsilon} r_1 \Big)^2}{2^k M_{\varepsilon} \Xi_{2, 1}^{2} } \Big) \\
& = \frac{ C_1 \Xi_{\theta, 1}^{\theta}}{ \Big( M_{\varepsilon} r_1^{\theta/(\theta-1) } \Big)^{\theta-1} } \sum_{k=1}^{\infty} \frac{2^k}{ 2^{\theta(k-1)} } 
    +  \sum_{k=1}^{\infty}  C_2 \exp\Big(- \frac{C_3 2^{k-2} M_{\varepsilon} r_1^2 }{ \Xi_{2, 1}^{2} } \Big),
    \end{split}
\end{equation*}
where $r_1 = r_1(d_*, \varepsilon) := \rho d_* - (1 + \rho) \varepsilon$.
For the third inequality, note that Condition~\ref{cond:41} holds. Therefore, we can apply Theorem~2 in \cite{WuWu2016}, and hence, there exist positive constants $C_1, C_2$ and $C_3$, such that the third inequality holds.

Recall the upper bound conditon on $d_*$, by which we have that there exists $K > \rho$ with
\begin{equation*}
\min_{t \geq \tau} (\mu_t - \mu_1) \geq d > K d_*.
\end{equation*}
Thus, for $\varepsilon > 0$ small enough such that $(1 + \rho) \varepsilon - (K-\rho) d_* < 0$, we have
\begin{equation*}
    \begin{split}
\eqref{eq:P2} 
& = \IP\left(\max _{\ell \geq M_{\varepsilon}}
\sum_{t=\tau}^{\tau+\ell-1}(- Z_t) + \ell (\hat\mu_1 - \mu_1 + \rho \hat d) - \sum_{t=\tau}^{\tau+\ell-1} (\mu_t - \mu_1)  \geq 0\right) \\
& \leq \IP\left(\max _{\ell \geq M_{\varepsilon}}
\sum_{t=\tau}^{\tau+\ell-1}(- Z_t) + \ell (1 + \rho) \varepsilon - \sum_{t=\tau}^{\tau+\ell-1} \Big( (\mu_t - \mu_1) - \rho d_* \Big) \geq 0\right) \\
& \qquad\qquad + \IP\Big( | \hat d - d_* | > \varepsilon \Big) + \IP\Big( | \hat\mu_1 - \mu_1 | > \varepsilon \Big).
    \end{split}
\end{equation*}
Further, for the first term in the previous bound, we have
\begin{equation*}
    \begin{split}
    & \IP\left(\max _{\ell \geq M_{\varepsilon}}
\sum_{t=\tau}^{\tau+\ell-1}(- Z_t) + \ell (1 + \rho) \varepsilon - \sum_{t=\tau}^{\tau+\ell-1} \Big( (\mu_t - \mu_1) - \rho d_* \Big) \geq 0\right) \\
& \leq \sum_{k=1}^{\infty} \IP\left(\max _{2^{k-1}M_{\varepsilon} \leq \ell \leq 2^k M_{\varepsilon}}
\sum_{t=\tau}^{\tau+\ell-1}(- Z_t) + \ell (1 + \rho) \varepsilon - \sum_{t=\tau}^{\tau+\ell-1} \Big( (\mu_t - \mu_1) - \rho d_* \Big) \geq 0\right) \\
& \leq \sum_{k=1}^{\infty} \IP\left(\max _{1 \leq \ell \leq 2^k M_{\varepsilon}}
\sum_{t=\tau}^{\tau+\ell-1}(- Z_t) \geq 2^{k-1} M_{\varepsilon} \Big( (K-\rho) d_* - (1 + \rho) \varepsilon \Big) \right) \\
& \leq \frac{ C_1 \Xi_{\theta, 1}^{\theta}}{ \Big( M_{\varepsilon} r_2^{\theta/(\theta-1) } \Big)^{\theta-1} } \sum_{k=1}^{\infty} \frac{2^k}{ 2^{\theta(k-1)} } 
    +  \sum_{k=1}^{\infty}  C_2 \exp\Big(- \frac{C_3 2^{k-2} M_{\varepsilon} r_2^2 }{ \Xi_{2, 1}^{2} } \Big),
    \end{split}
\end{equation*}
where $r_2 = r_2(d_*, \varepsilon) := (K-\rho) d_* - (1 + \rho) \varepsilon$ and we have applied Theorem~2 from \cite{WuWu2016} again to obtain the third inequality.
For $c$ with $0 < c < \min\{\rho, K-\rho\} /(1+\rho)$ choose $\varepsilon = c d_*$. Then, let $C := \min\{\rho, K-\rho\} /(1+\rho) - c > 0$ and
$M_{\varepsilon} = \tilde M_{\varepsilon} d_*^{-\theta/(\theta-1)}$.
This yields
\begin{equation}\label{eqn:bnd_err_hattau}
\begin{split}
    & \IP\left(\left|\hat{\tau}_n - \tau\right| \geq \tilde M_{\varepsilon} d_*^{-\theta/(\theta-1)} \right) \\
    & \leq 2 \frac{ C_1 \Xi_{\theta, 1}^{\theta}}{ \Big( \tilde M_{\varepsilon} C^{\theta/(\theta-1) } \Big)^{\theta-1} } \sum_{k=1}^{\infty} \frac{2^k}{ 2^{\theta(k-1)} } 
    +  2 \sum_{k=1}^{\infty}  C_2 \exp\Big(- \frac{C_3 2^{k-2} \tilde M_{\varepsilon} C^2 }{ \Xi_{2, 1}^{2} } \Big) + o(1),
\end{split}
\end{equation}
where the $o(1)$ follows from Lemmas~\ref{lem:mu1hat} and~\ref{lem:dhat} and the fact that $(\log(m)/k)^{1/2} = o(d_*)$ and $(k \eta)^{-1/2} = o(d_*)$, which can be seen from
\[0 \leq \frac{1}{k \eta d_*^2} \leq \frac{\log(m)}{k d_*^2} \leq \frac{\log(m)}{k d^2} = o(1),\]
where the first inequality holds for $\log(m) \eta \geq 1$, the second inequality is due to $d_* \geq d$ and the $o(1)$ is condition (C3).
Noting that the bound in~\eqref{eqn:bnd_err_hattau} will be arbitrarily small for all $n$ large enough, if $\tilde M_{\varepsilon}$ is chosen large enough, which yields the bound of the theorem.
\hfill\qed

\noindent
\subsection{Proof of the Lemmas in Sections~\ref{sec:proofs:thm:sigmahat}--\ref{sec:proofs:thm:tauhat}}
\label{sec:proofs:auxresults}

\noindent
\subsubsection{Proof of Lemma~\ref{lem:hatL}}
\label{sec:proofs:hatL}

It suffices to prove the assertion for
\begin{equation}
    \label{eqn:hatL:cond_tn}
    t_n \in \mathbb{N}, \quad \text{with} \quad
    t_n = o(\eta) \quad \text{and} \quad
    t_n \to \infty.
\end{equation}
Otherwise, say $0 \leq t'_n = o(\eta)$ is given, it is always possible to choose $t_n$ satisfying~\eqref{eqn:hatL:cond_tn} and $t'_n \leq t_n$, such that, if the assertion is proven for $t_n$ satsifying~\eqref{eqn:hatL:cond_tn}, it holds also with $t'_n$, since
\begin{equation*}
1 \geq \IP(t'_n \leq \hat{L} \leq \eta - t'_n) \geq \IP(t_n \leq \hat{L} \leq \eta - t_n) \to 1.
\end{equation*}

Denote the order statistics of $R_1, \ldots, R_s$, $s = 1,\ldots,m$, by
\begin{equation}\label{eqn:ord_stat_R}
R^{(j)}_s := \min\Big\{ x \in \R : \sum_{i=1}^s I\{R_i \leq x\} \geq j\Big\}, \quad j=1,\ldots,s.
\end{equation}
Then, the generalized version of the index of $\hat L$ is $\hat L_J := \max\{i : R_i \leq R^{(J)}_m\}$ and the version of $\hat L$ defined in~(7) is $\hat L_1$.
For any $x \in \R$ and $s = 1, \ldots, m-1$, we have
\begin{equation*}
\begin{split}
    \Big\{ \min_{j \geq s+1} R_j \geq x \Big\} \cap \Big\{ R^{(J)}_s < x \Big\} \subset \{\hat L_J \leq s\} 
\end{split}
\end{equation*}
and, for $s = 2, \ldots, m$, we have
\begin{equation*}
\begin{split}
    \Big\{ \min_{j \leq s-1} R_j \geq x \Big\} \cap \Big\{ \min_{j \geq s} R_j < x \Big\} \subset \{\hat L_1 \geq s\}.
\end{split}
\end{equation*}

Employing $A \cap B \subset C \Leftrightarrow C^c \subset A^c \cup B^c$ for events $A, B, C$ with complements $A^c, B^c, C^c$, this implies that for any $x_n, x'_n \in \mathbb{R}$
\begin{equation*}
\begin{split}
    & \IP(t_n \leq \hat{L}_J \leq \eta - t_n)
    \geq \IP( \{ t_n \leq \hat{L}_1 \} \cap \{\hat{L}_J \leq \eta - t_n \} ) \\
    & = 1 - \IP(\{ t_n > \hat{L}_1 \} \cup \{ \hat{L}_J > \eta - t_n \})
    \geq 1 - \IP(\hat{L}_1 < t_n ) - \IP( \hat{L}_J > \eta - t_n ) \\
    & \geq 1 - \IP\Big( \min_{j \leq t_n-1} \frac{\sqrt{k} (R_j - \mu_1)}{\sigma_{\infty}} < x_n \Big) - \IP\Big( \min_{j \geq t_n} \frac{\sqrt{k} (R_j - \mu_1)}{\sigma_{\infty}} \geq x_n \Big) \\
    & \qquad - \IP\Big( \min_{j \geq \eta - t_n+1} \frac{\sqrt{k} (R_j - \mu_1)}{\sigma_{\infty}} < x'_n \Big) - \IP\Big( \frac{\sqrt{k} (R^{(J)}_{\eta - t_n} - \mu_1)}{\sigma_{\infty}} \geq x'_n \Big). \\
\end{split}
\end{equation*}
Next, denote $\tilde R_j := \sqrt{k} (R_j - \IE R_j)/\sigma_{\infty}$ and observe that $R_j - \mu_1 = R_j - \IE R_j$ for $j \leq \eta$, $R_j - \mu_1 \geq R_j - \IE R_j$ for $j > \eta$, and $R_j - \mu_1 \geq R_j - \IE R_j + d$ for $j > \eta+1$. Thus, for $n$ large enough such that $t_n < \eta$, we have
\begin{equation}\label{eqn:bnd_P_hatL}
\begin{split}
    \IP(t_n \leq \hat{L} \leq \eta - t_n)
    &  \geq 1 - \IP\Big( \min_{j=1,\ldots,t_n} \tilde R_j < x_n \Big)
    - \IP\Big( \min_{j=t_n,\ldots, \eta} \tilde R_j \geq x_n \Big) \\
    &  \qquad - \IP\Big( \min_{\eta - t_n+1 \leq j \leq \eta + 1} \tilde R_j < x'_n \Big) - \IP\Big( \tilde R^{(J)}_{\eta - t_n} \geq x'_n \Big) \\
    & \qquad - \IP\Big( \min_{j=\eta+2, \ldots, m} \tilde R_j + \sqrt{k} d/\sigma_{\infty} < x'_n \Big) I\{ \eta + 1 < m \},
\end{split}
\end{equation}
where $\tilde R^{(J)}_{\eta-t_n}$ denotes the $J$th order statistic of $\tilde R_1, \ldots, \tilde R_{\eta - t_n}$; cf. \eqref{eqn:ord_stat_R}. We have also employed that for sets $A \subset B$, we have $\min_{j \in B} \tilde R_j \leq \min_{j \in A} \tilde R_j$.
Next, let $(Z_i^c)_{i \in \mathbb Z}$ and $\mathbb{B}_c (\cdot)$ be as in~(19).
Denote, for $i \in \mathbb{N}$,
\begin{equation*}
\begin{split}
    W_{i} & = k^{-1/2} ( \mathbb{B}_c (ik) - \mathbb{B}_c \big( (i-1)k \big), \text{ and} \\
    M_{i} & = k^{-1/2} \sum_{j=(i-1)k+1}^{ik} \left( Z_j^c / \sigma_{\infty} - \left(\mathbb{B}_c (j) - \mathbb{B}_c (j-1) \right) \right).
\end{split}
\end{equation*}
Further, denoting $R_i^c := k^{-1} \sum_{j=(i-1)k+1}^{ik} ( Z_j^c + \mu_j )$, we have (jointly for all $i$)
\begin{equation*}
\tilde R_i \stackrel{\mathcal{D}}{=} \tilde R_i^c := \sqrt{k} \frac{R_i^c - \IE R_i^c}{\sigma_{\infty}} = W_{i} + M_{i}.
\end{equation*}
Note that $W_i$ are i.\,i.\,d.\ standard normally distributed and that by Corollary 2.1 in \cite{BerkesLiuWu2014}, under Condition~3.1, we have
\begin{equation}\label{eqn:KMT_M}
    M^{(m)} := \max_{i =1, \ldots m} | M_{i} | = \textit{o}_{a.s.} (n^{1/\theta} k^{-1/2}).
\end{equation}
With the above notation we have
\[(\min_{i \in \mathcal{I}} W_i) - M^{(m)} \leq \min_{i \in \mathcal{I}} (W_i + M_i) \leq (\min_{i \in \mathcal{I}} W_i) + M^{(m)}, \text{ for any $\mathcal{I} \subset \{1, \ldots, m\}$,}\]
and denoting by $W^{(J)}_{\eta - t_n}$ the $J$th order statistic of $W_1, \ldots, W_{\eta - t_n}$, we have
\[W^{(J)}_{\eta - t_n} - M^{(m)} \leq \tilde R^{(J)}_{\eta - t_n} \leq W^{(J)}_{\eta - t_n} + M^{(m)}.\]

Denote the cdf of the Gumbel distribution by $G(x) := \exp(-\exp(-x))$ and the scaling factor
\[\gamma_t = (2 \log t-\log \log t-\log (4 \pi))^{1 / 2}
= (2 \log t)^{1 / 2} - \frac{\log(4\pi \log(t))}{2 (2 \log t)^{1 / 2}} + o\big( (\log t)^{-1 / 2} \big),\]
as $t \to \infty$.
From Fisher–Tippett–Gnedenko theorem we have that, for any sequence $N \rightarrow \infty$,
\begin{equation}
\label{convToGumbel}
\sup_x \Big| \IP\Big( \gamma_N \Big( \max_{j=1,\ldots,N} W_j - \gamma_N \Big) \leq x \Big) - G(x) \Big| = o(1).
\end{equation}
Further, by Theorem 2.2.2 in \cite{LeadbetterEtAl1983} together with the fact that $-W_N^{(J)}$ is the $J$th largest among $-W_1, \ldots, -W_N$ and the fact that $(-W_1, \ldots, -W_N)$ has the same distribution as $(W_1, \ldots, W_N)$, this implies that
\begin{equation}
\label{convOrderStat}
\sup_x \Big| \IP\Big( \gamma_N \Big( - W^{(J)}_N - \gamma_N \Big) \leq x \Big) - G(x) \sum_{s=0}^{J-1} \frac{(-\log G(x))^s}{s!} \Big| = o(1).
\end{equation}
Thus, for $x_n = -(\gamma_{t_n} + \gamma_{\eta - t_n + 1})/2$, we have
\begin{equation*}
\begin{split}
    & \IP\Big( \min_{j=1,\ldots,t_n} \tilde R_j < x_n \Big)
    \leq \IP\Big( \min_{j=1,\ldots,t_n} W_j < x_n + M^{(m)} \Big) \\
    & = \IP\Big( \gamma_{t_n} \Big(\Big( \max_{j=1,\ldots,t_n} -W_j\Big) - \gamma_{t_n} \Big) > \gamma_{t_n} ( (-\gamma_{t_n} + \gamma_{\eta - t_n + 1})/2 - M^{(m)}) \Big) \\
    & \leq \IP\Big( \Big\{ \gamma_{t_n} \Big(\Big( \max_{j=1,\ldots,t_n} -W_j\Big) - \gamma_{t_n} \Big) > \gamma_{t_n} (\gamma_{\eta - t_n + 1} -\gamma_{t_n}) ( 1/2 - M^{(m)}/ (\gamma_{\eta - t_n + 1} -\gamma_{t_n}) ) \Big\} \\
    & \quad \cap \Big\{M^{(m)}/ (\gamma_{\eta - t_n + 1} -\gamma_{t_n}) \leq 1/4 \Big\} \Big) + \IP\Big( M^{(m)}/ (\gamma_{\eta - t_n + 1} -\gamma_{t_n}) > 1/4 \Big) \\
    & \leq 1 - G\Big( \gamma_{t_n} (\gamma_{\eta - t_n + 1} -\gamma_{t_n})/4 \Big) + o(1) \to 0.
\end{split}
\end{equation*}
The $o(1)$ in the last line relates to two convergences: for the first probability, decrease the lower bound for the scaled maximum to $\gamma_{t_n} (\gamma_{\eta - t_n + 1} -\gamma_{t_n})/4$ and then use~\eqref{convToGumbel}; for the second probability to vanish, note the fact that $1/(\gamma_{\eta - t_n + 1} -\gamma_{t_n}) = o\big( (\log \eta)^{1/2} \big)$, as $n \to \infty$, which together with~\eqref{eqn:KMT_M} and (C1) implies that $M^{(m)}/ (\gamma_{\eta - t_n + 1} - \gamma_{t_n}) = o_{\rm a.\,s.}(1)$. For the final convergence we used the fact that $\gamma_{t_n} (\gamma_{\eta - t_n + 1} -\gamma_{t_n}) \to \infty$.
For the second probability on the right-hand side of~\eqref{eqn:bnd_P_hatL} we argue similarly and have
\begin{equation*}
\begin{split}
    & \IP\Big( \min_{j=t_n,\ldots, \eta} \tilde R_j \geq x_n \Big)
    \leq \IP\Big( \min_{j=t_n,\ldots, \eta} W_j \geq x_n - M^{(m)} \Big) \\
    & = \IP\Big( \gamma_{\eta - t_n + 1} \Big( \Big( \max_{j=t_n,\ldots, \eta} -W_j \Big) - \gamma_{\eta - t_n + 1} \Big) \\
    & \qquad\qquad \leq - \gamma_{\eta - t_n + 1} ((\gamma_{\eta - t_n + 1} - \gamma_{t_n})/2 - M^{(m)}) \Big) \\
    & \leq G\Big( - \gamma_{\eta - t_n + 1} (\gamma_{\eta - t_n + 1} - \gamma_{t_n})/4 \Big) + o(1) \to 0, \\
\end{split}
\end{equation*}
where we have $- \gamma_{\eta - t_n + 1} (\gamma_{\eta - t_n + 1} - \gamma_{t_n})/4 \to -\infty$ due to condition (C1).

The probabilities $\IP\big( \min_{\eta - t_n+1 \leq j \leq \eta + 1} \tilde R_j < x'_n \big)$ and $\IP\big( \tilde R^{(J)}_{\eta - t_n} \geq x'_n \big)$ are treated analogously, with $x'_n = -(\gamma_{t_n + 1} + \gamma_{\eta - t_n})/2$ and~\eqref{convOrderStat} instead of~\eqref{convToGumbel} for the second probability that involves the $J$th order statistic. It remains to bound the fifth one:
\begin{equation}\label{eqn:bnd_term5}
\begin{split}
    & \IP\Big( \min_{j=\eta+2, \ldots, m} \tilde R_j + \sqrt{k} d/\sigma_{\infty} < x'_n \Big)
    \leq (m - \eta - 1) \IP\Big( | \sum_{i=1}^k Z_i| > k d + \sigma_{\infty} |x'_n| \sqrt{k}  \Big) \\
    & \leq 
    (m - \eta - 1) \frac{C_{1} \Xi_{\theta, 1}^{\theta} k}{(k d + \sigma_{\infty} |x'_n| \sqrt{k})^{\theta}}+ C_2 (m - \eta - 1) \exp \left( - \frac{C_3 (k d + \sigma_{\infty} |x'_n| \sqrt{k})^2 }{k \Xi_{2, 1}^{2}}  \right),
\end{split}
\end{equation}
where we used sub-additivity of $\IP$ and the fact that $-x'_n = |x'_n|$ for the first inequality and applied Theorem~2 in~\cite{WuWu2016} for the second inequality. Now, we see that the first term on the right-hand side of~\eqref{eqn:bnd_term5} is $o(1)$ by employing the fact that $m - (\eta+1) \leq (n-\tau)/k$ and
\[\frac{n - \tau}{(k d + \sigma_{\infty} |x'_n| \sqrt{k})^{\theta}}
= \mathcal{O}\Big(
\frac{n }{ (k d + (k \log \eta)^{1/2} )^{\theta}}\Big),
\]
due to $|x'_n| \asymp (\log \eta)^{1/2}$. So the first term vanishes due to condition (C2). For the second term on the right-hand side of~\eqref{eqn:bnd_term5} note that $0 \leq \log(m - \eta - 1) \leq \log(m) \ll k d^2 \to \infty$, due to condition~(C3).\hfill\qed

\noindent
\subsubsection{Proof of Lemma~\ref{lem:smSpecDens}}
\label{sec:proofs:smSpecDens}

Denote $\xi_t := (\varepsilon_i : i \leq t)$ and define the projection operator $\mathcal{P}_t(Z) := \IE(Z | \xi_t) - \IE(Z | \xi_{t-1})$, $Z \in \mathcal{L}^2$. Then, for $k \ge 0$, we have
\begin{equation*}
\begin{split}
    |\gamma(k)| &  = |\gamma(-k)|
    = |\IE( Z_k Z_0 )|
    = \Big| \IE\Bigg( \Big( \sum_{i \leq 0} \mathcal{P}_i Z_k \Big) \Big( \sum_{i \leq 0} \mathcal{P}_i Z_0 \Big) \Bigg) \Big| \\
    & = \Bigg| \sum_{i \leq 0} \IE\Bigg( \Big( \mathcal{P}_i Z_k \Big) \Big( \mathcal{P}_i Z_0 \Big) \Bigg) \Bigg|
    \leq \sum_{i \leq 0} \| \mathcal{P}_i Z_k \|_2 \| \mathcal{P}_i Z_0 \|_2
    \leq \sum_{i \geq 0} \delta_{i+k, 2} \delta_{i, 2}
\end{split}
\end{equation*}
This implies $\sum_{k = -\infty}^{\infty} |\gamma(k)| \leq \Theta_{0,2}^2$ and, since $\sum_{k=1}^{\infty} k \delta_{k,2} \le \sum_{i=0}^{\infty} \Theta_{i,2}$,
\begin{equation*}
\begin{split}
    & \sum_{k = -\infty}^{\infty} |k \gamma(k)| \leq 2 \sum_{k = 1}^{\infty} k \sum_{i \geq 0} \delta_{i+k, 2} \delta_{i, 2} \\
    & \quad = 2  \sum_{i = 0}^{\infty} \delta_{i, 2} \sum_{k = 1}^{\infty} k \delta_{i+k, 2}
    \leq 2  \Big( \sum_{i = 0}^{\infty} \delta_{i, 2} \Big) \Big( \sum_{k = 1}^{\infty} k \delta_{k, 2} \Big)
    \le 2 \Theta_{0,2} \sum_{i=0}^{\infty} \Theta_{i,2}.
\end{split}
\end{equation*}
Elementary calculations show that $\gamma_{\theta} > 1$ if $\theta > 4$. Thus $\sum_{i=0}^{\infty} \Theta_{i,2} < \infty$ holds for all $\theta > 2$.
\hfill\qed

\noindent
\subsubsection{Proof of Lemma~\ref{lem:rate_|hat_mu1-mu1|}}
\label{sec:proofs:rate_|hat_mu1-mu1|}

To show~\eqref{eq:rate_|hat_mu1-mu1|}, we prove an equivalent proposition:
\begin{equation}
\label{eq:rate_|hat_mu1-mu1|_equiv}
|\hat{\mu}_0 - \mu_1| = \mathcal{O}_{\IP}\Big(\frac{a_n}{\sqrt{\eta k}}\Big)
\end{equation}
for any $\{a_n\}$, $a_n \to \infty$. 
Choose $r_n = a_n^{-1}$, then,
\begin{align*}
& \IP \Big( |\hat{\mu}_0 - \mu_1| \geq \frac{a_n}{\sqrt{\eta k}} M   \Big) \\
& \leq \IP ( \hat{L} < \eta r_n) + \IP \left(\eta \geq \hat{L} \geq \eta r_n, \frac{|\sum_{i=1}^{\hat{L} k} (X_i - \mu_1)|}{\hat{L} k} \geq \frac{a_n}{\sqrt{\eta k}} M \right) + \IP ( \hat{L} > \eta)  \\
& \leq 1 - \IP \big( \lceil \eta r_n \rceil  \leq \hat{L} \leq \eta - \lceil \eta r_n \rceil \big) + \IP \left( \max_{j \geq \eta r_n k} \Big| \frac{1}{j} \sum_{i=1}^{j} Z_i\Big| \geq \frac{a_n}{\sqrt{\eta k}}  M \right) = o(1),
\end{align*}
which implies~\eqref{eq:rate_|hat_mu1-mu1|_equiv}.

The $o(1)$ in the above follows from applying Lemma~\ref{lem:hatL} with $t_n := \lceil \eta r_n \rceil = o(\eta)$ to the first probability and Lemma~\ref{lem:maxSum_gn} below with $g_n^{-1/2} = (\eta k r_n)^{-1/2} = a_n^{1/2} (\eta k)^{-1/2}$ to the second probability.

\begin{lemma}
    \label{lem:maxSum_gn}
    Grant condition~\ref{cond:41} with $\alpha = 1$. Then, for any sequence $g_n \in \mathbb{N}, g_n \to \infty$, we have 
    \begin{equation}
        \label{eq:rate_max|S_j/j|}
        \max_{j \geq g_n} \left| \frac{1}{j} \sum_{i=1}^j Z_i \right| = \mathcal{O}_{\IP} \Big( \frac{1}{\sqrt{g_n}} \Big).
    \end{equation}
\end{lemma}

\begin{proof}
Now, for any $G > 0$, note that by Bonferroni's inequality, we have 
\begin{equation*}
\begin{split}
\IP \left ( \max_{j \geq g_n} \left| \frac{1}{j} \sum_{i=1}^j Z_i \right| \geq \frac{G}{\sqrt{g_n}}\right)
 & \leq \sum_{k=1}^{\infty} \IP \left ( \max_{2^{k-1} g_n \leq j \leq 2^k g_n} \left| \frac{1}{j} \sum_{i=1}^j Z_i \right| \geq \frac{G}{\sqrt{g_n}} \right) \\
 & \leq \sum_{k=1}^{\infty} \IP \left ( \max_{2^{k-1} g_n \leq j \leq 2^k g_n} \left| \frac{1}{2^{k-1} g_n } \sum_{i=1}^j Z_i \right| \geq \frac{G}{\sqrt{g_n}} \right) \\
 & \leq \sum_{k=1}^{\infty} \IP \left ( \max_{1 \leq j \leq 2^k g_n} \left| \sum_{i=1}^j Z_i  \right| \geq 2^{k-1} G \sqrt{g_n} \right).
\end{split}
\end{equation*}

As argued before, we have that Condition~\ref{cond:41} holds, we can apply the Nagaev-type inequality under dependence from Theorem~2 in \cite{WuWu2016}, and hence, there exists positive constants $C_1$, $C_2$ and $C_3$, such that
\begin{equation*}
\begin{split}
& \IP \left ( \max_{1 \leq j \leq 2^k g_n} \left| \sum_{i=1}^j Z_i  \right| \geq 2^{k} G \sqrt{g_n}/2 \right) \\
& \quad \leq \frac{C_{1} \Xi_{\theta, 1}^{\theta} (2^k g_n)}{\left(2^{k} G \sqrt{g_n}/2\right)^{\theta}}+ C_2 \exp \left( - \frac{C_3 2^{2k} G^2 g_n/4 }{2^k g_n \Xi_{2, 1}^{2}}  \right) \\
& \quad = C_{1} \Xi_{\theta, 1}^{\theta} \frac{ 1}{2^{k(\theta-1)}} \frac{ 1 }{ g_n^{\theta/2-1} (G/2)^{\theta} }+ C_2 \exp \left( - \frac{C_3}{\Xi_{2, 1}^{2}}  2^{k} (G/2)^2  \right).
\end{split}
\end{equation*}
Thus, employing $k \leq 2^k$, we have
\begin{align*}
\IP \left ( \max_{j \geq g_n} \left| \frac{1}{j} \sum_{i=1}^j Z_i \right| \geq G/\sqrt{g_n} \right)
& \leq C_{1} \Xi_{\theta, 1}^{\theta} \frac{ 2^{\theta-1} }{1 - 2^{\theta-1}} \frac{ 1 }{ g_n^{\theta/2-1} (G/2)^{\theta} }+ \frac{C_2}{\exp ( \frac{C_3}{4 \Xi_{2, 1}^{2}} G^2 ) - 1},
\end{align*}
which is arbitrarly small for $G$ large enough.
The result follows.
\end{proof}

\subsubsection{Proof of Theorem~\ref{thm:etahat}}
\label{sec:proofs:thm:etahat}

\noindent
We use approximations to $\hat D_j$ and $R_{j}$ defined by
\begin{equation}
\label{def:tildeDandR}
\tilde{D}_j := \sqrt{k} (\tilde{R}_{j} - \mu_1) / \sigma_{\infty}, \quad \tilde{R}_{j} := \E (R_{j} | \varepsilon_{(j-1)k + 1}, \ldots, \varepsilon_{j k}), \quad j = 1,2,\ldots.
\end{equation}
Note that the $\tilde R_{1}, \ldots, \tilde R_{m}$ are independent, because the $\varepsilon_t$ are independent, and hence the $\tilde{D}_j$ are independent. In the proof of Theorem~\ref{thm:etahat}, we will use the following result that asserts a rate for the approximation of the $R_{j}$ by the independent $\tilde{R}_{j}$.
\begin{lemma}
	\label{lem:rate_Delta_s}
	Assume Condition~3.1 holds. Then, we have
	\begin{equation}
	\label{eq:rate_Delta_s}
	\max_{j=1,\ldots, m} \big| R_{j} - \tilde{R}_{j} \big| = O_\IP\Big( \frac{m^{1/\theta}}{k} \Big), \quad n \to \infty.
	\end{equation}
\end{lemma}
The proof of Lemma~\ref{lem:rate_Delta_s} is deferred to Section~\ref{sec:proofs:rate_Delta_s}.
Based on the $\tilde{D}_j$ we will also define approximations $\tilde{I}_j$ to the test decisions $I_j$. Again, the $\tilde{I}_j$ will be independent and we will be able to apply the following result, which we think is interesting in itself.

\begin{lemma}\label{lem:sum_Ber}
	Let $I_1, I_2, \ldots$ be a sequence of independent Bernoulli-distributed random variables with $\IP(I_i = 1) = p_i = 1 - \IP(I_i = 0)$.
	Then, we have
	\[\IP\Big( \max_{j \geq 1} \sum_{i=1}^j (2 I_i - 1) \geq 0 \Big)
	= \sum_{k=1}^{\infty} \sum_{(i_1, \ldots, i_{2k}) \in A_k} \prod_{\ell=1}^{2k} \IP(2 I_\ell - 1 = i_\ell ),
	\]
	where
	\[A_k = \big\{(i_1, \ldots, i_{2k}) \in \{-1,1\}^{2k} : \sum_{\ell=1}^{2k} i_{\ell} = 0 \text{ and } \sum_{\ell=1}^j i_{\ell} \leq 0, \forall j=1,\ldots,2k\big\}.\]
\end{lemma}
The proof of Lemma~\ref{lem:sum_Ber} is deferred to Section~\ref{sec:proofs:sum_Ber}.

Now we proceed with the proof of Theorem~\ref{thm:etahat}.
Due to the assumed conditions (C1), (C2) and (C3), we can apply Lemma~\ref{lem:rate_|hat_mu1-mu1|}. Due to Condition~\ref{cond:41}, we can apply Lemma~\ref{lem:rate_Delta_s}.
Our aim is to prove 
\begin{equation}
\label{eqn:eta_Op1}
\IP\left(\left|\hat\eta - \eta\right| \geq 1  \right) \to 0, \text{ as $n \to \infty$}.
\end{equation}
Note that
\begin{equation*}
\begin{split}
\Omega & = \left(\left\{\sum_{j=\hat\eta+1}^{\eta}\left(2 \hat I_{j}-1\right) \geq 0\right\} \cap\left\{\hat\eta<\eta\right\}\right) \\
& \qquad \cup \left(\left\{\sum_{j=\eta+1}^{\hat\eta}\left(2 \hat I_{j}-1\right) \leq 0\right\} \cap\left\{\hat\eta>\eta\right\}\right)
\cup \left\{\hat\eta=\eta\right\}.
\end{split}
\end{equation*}
Thus,
\begin{equation*}
\begin{split}
\IP\left(\left|\hat\eta-\eta\right| \geq 1 \right)
& = \IP\left(\max _{\ell = 1, \ldots, \eta-1} \sum_{j=\eta-\ell+1}^{\eta}\left(2 \hat I_{j}-1\right) \geq 0\right) \\
& \qquad +\IP\left(\max _{\ell = 1, \ldots, m - \eta-1} \sum_{j=\eta+1}^{\eta+\ell}\left(1-2 \hat I_{j}\right) \geq 0\right).
\end{split}
\end{equation*}

Fix $c_1, c_2 > 0$ and $\lambda \in (0,1)$ as in (C5); recall the notation defined in~\eqref{def:tildeDandR}, then we have
\begin{align*}
& \IP \left(  \max_{\ell = 1, \ldots, \eta-1} \sum_{j = \eta -\ell+1}^{\eta} (2 \hat I_j -1) \geq 0   \right )
\leq \IP \Big(  \max_{\ell = 1, \ldots, \eta-1} \sum_{j = \eta -\ell+1}^{\eta} (2 \hat I_j -1) \geq 0, \\
& \qquad\qquad |\hat\mu_0 - \mu_1|< \frac{c_1}{\sqrt{k}}, \forall j > 1, \ | R_{j} - \tilde{R}_{j} | < \frac{c_2}{\sqrt{k}}, \ (1-\lambda)^{1/2} \leq \frac{\hat\sigma_{\infty}}{\sigma_{\infty}} \leq (1+\lambda)^{1/2} \Big) \\
& \quad + \IP\Big(|\hat\mu_0 - \mu_1| > \frac{c_1}{\sqrt{k}}\Big) + \IP\Big( \max_{j=1,\ldots,m} | R_{j} - \tilde{R}_{j} | > \frac{c_2}{\sqrt{k}} \Big) + \IP\Big( | \hat\sigma_{\infty}^2 - \sigma_{\infty}^2 | >   \lambda \sigma_{\infty}^2 \Big) \\
& \quad =: A_n + B_n + C_n + D_n.
\end{align*}

By Lemma~\ref{lem:rate_|hat_mu1-mu1|} and the fact that $(\eta k)^{-1/2} \ll k^{-1/2}$, by condition (C0), we have $B_n \to 0$. By Lemma~\ref{lem:rate_Delta_s} and condition~(C4) we have $C_n \to 0$. By condition (C5) we have $D_n \to 0$, as $n \to \infty$.
Now define
\[
\tilde{I}_j = \mathbf{1}\Big\{\tilde{D}_j \geq \frac{z_{1 - 1 / m}}{(1-\lambda)^{1/2}} - \frac{c_1 + c_2}{\sigma_{\infty}}\Big\}, \quad j=1,2,\ldots \, .
\]
Note that $\tilde{I}_1, \tilde{I}_2, \ldots, \tilde{I}_{\eta}$ are i.\,i.\,d.\ Bernoulli-distributed with $\IP(\tilde{I}_1 = 1) =: p \to 0$, as $n \to \infty$, due to $m \to \infty$ and $\tilde D_1 = \mathcal{O}_{\IP}(1)$.
Let $\tilde{I}_{\eta+1}, \ldots, \tilde{I}_{\eta+2}, \ldots$ be independent and distributed as~$\tilde{I}_{1}$. Then, Lemma~\ref{lem:sum_Ber} entails that
\[\IP\Big( \max_{j \geq 1} \sum_{i=1}^j (2 \tilde I_i - 1) \geq 0 \Big)
= \sum_{k=1}^{\infty} |A_k| \big( p (1-p) \big)^k
\leq \min\Big\{ \frac{4 p}{1-4 p}, 1 \Big\} \leq  8 p,
\]
where we have used the fact that $|A_k| \leq 2^{2k}$ and $p (1-p) \leq p$.
On 
\[\{|\hat\mu_0 - \mu_1|< c_1/\sqrt{k}\} \cap \{| R_{j} - \tilde{R}_{j} | < c_2/\sqrt{k} \} \cap \Big\{ (1-\lambda)^{1/2} \leq \frac{\hat\sigma_{\infty}}{\sigma_{\infty}} \leq (1+\lambda)^{1/2} \Big\},\] we have $\Tilde{I}_j \geq \hat I_j$. Thus,
\begin{equation*}
A_n  \leq \IP \left(  \max_{\ell = 1, \ldots, \eta-1} \sum_{j=\eta - \ell+1}^{\eta} (2\Tilde{I}_j -1) \geq 0    \right)
\leq \IP\Big( \max_{j \geq 1} \sum_{i=1}^j (2 \tilde I_i - 1) \geq 0 \Big) \to 0.
\end{equation*}

Similarly, using the notation $ \hat J_j = 1 - \hat I_j$ and
\begin{equation*}
\begin{split}
\tilde{J}_j & := \mathbf{1}\Big\{\tilde{D}_j < \frac{z_{1 - 1 / m}}{(1+\lambda)^{1/2}} + \frac{c_1 + c_2}{\sigma_{\infty}}\Big\}, \quad j = \eta + 1, \ldots, m
\end{split}
\end{equation*}
and adding independent $\tilde J_1, \ldots, \tilde J_{\eta}$ and $\tilde J_{m + 1}, \tilde J_{m + 2}, \ldots$ that have distribution as $J_{m+1}$, we have that
\begin{equation*}
\bar p := \sup_{\substack{j \geq 1 \\ j \neq \eta+1}} \IP(\tilde J_j = 1) \leq \IP\Big( \tilde D_j - \IE \tilde D_j \leq \frac{z_{1 - 1 / m}}{(1+\lambda)^{1/2}} - ( \sqrt{k} d + c_1 + c_2) / \sigma_{\infty} \Big) \to 0,
\end{equation*}
as $n \to \infty$, because $\tilde{D}_j - \IE \tilde{D}_j$ are i.\,i.\,d.\ and
\[\frac{z_{1 - 1 / m}}{(1+\lambda)^{1/2}} - \E \tilde{D}_j \leq \frac{z_{1 - 1 / m}}{(1+\lambda)^{1/2}} - \sqrt{k} d \rightarrow -\infty,\] for $j \neq \eta + 1$. The right hand side diverges, as $z_{1 - 1 / m} \asymp (2 \log(m))^{1/2} \ll \sqrt{k} d$, where the $\ll$ is condition (C3).
Noting that $1 - \hat I_j \leq \tilde J_j$, we see that we have
\begin{equation*}
\begin{split}
\IP\left(\max _{\ell = 1, \ldots, m - \eta-1} \sum_{j=\eta+1}^{\eta+\ell}\left(1-2 I_{j}\right) \geq 0\right)
& \leq \IP\Big( \max_{j \geq 1} \sum_{i=1}^j (2 \tilde J_i - 1) \geq 0 \Big) \\
& \leq \min\Big\{ \sum_{k=1}^{\lfloor \eta / 2 \rfloor} (4 \bar p)^k + \sum_{k=\lfloor \eta / 2 \rfloor + 1}^{\infty} 4^k \bar p^{k-1}, 1 \Big\} \rightarrow 0.
\end{split}
\end{equation*}

This concludes the proof of Theorem~\ref{thm:etahat}.
 \hfill\qed

\noindent
\subsubsection{Proof of Lemma~\ref{lem:rate_Delta_s}}
\label{sec:proofs:rate_Delta_s}

By the sub-additivity of $\IP$ and the stationarity of $R_{j} - \tilde{R}_{j}$, we have for $u > 0$, that
$$
\IP \Big(\max_{j = 1, \ldots, m} \left | R_{j} - \tilde{R}_{j} \right | > u \Big) \leq m \cdot \IP \left( | R_{1} - \tilde{R}_{1}  | > u \right ).
$$

Let $\Delta_j := \E (R_{1} | \varepsilon_k, \ldots, \varepsilon_j) - \E (R_{1} | \varepsilon_k, \ldots, \varepsilon_{j-1})$, then it follows that 
$$
\tilde{R}_{1} - R_{1}  = \sum_{j=-\infty}^{1} \Delta_j.
$$
Applying Markov's inequality, we have
$$
\IP \Big( \Big| \sum_{j=-\infty}^{1} \Delta_j \Big| > u \Big) \leq \frac{\E |\sum_{j=-\infty}^{1} \Delta_j|^{\theta} }{u^{\theta}} = \frac{1 }{u^{\theta}} \Big\| \sum_{j=-\infty}^{1} \Delta_j \Big\|^{\theta}_{\theta}.
$$

Since $\{\Delta_j\}$ is a martingale difference sequence, we may apply the Burkholder-Davis-Gundy inequality to obtain a bound for the right-hand side of the previous inequality
\[
\Big\| \sum_{j=-\infty}^{1} \Delta_j \Big\|_{\theta}^2 \leq c_{\theta} \sum_{j=-\infty}^{1} \|\Delta_j \|_{\theta}^2,
\]
where the constant $c_{\theta}$ only depends on $\theta$.
Recall that $R_{1} = \frac{1}{k} \sum_{i = 1}^{k} X_i$. Therefore, by the triangle inequality and Jensen's inequality, we have
\begin{equation*}
\begin{split}
\|\Delta_j \|_{{\theta}} & \leq \frac{1}{k} \sum_{i=1}^k  \left(\|\E (Z_i | \varepsilon_i, \ldots, \varepsilon_j) - \E (Z_i | \varepsilon_i, \ldots, \varepsilon_{j-1}) \|_{{\theta}} \right) \\
& \leq \frac{1}{k} \sum_{i=1}^k \delta_{(i-j +1),{\theta}} \leq \frac{1}{k}  \Theta_{2-j,{\theta}}.
\end{split}
\end{equation*}
Combining the results above, it follows that
\[
\IP \Big(\max_{j=1,\ldots,m} | R_{j} - \tilde{R}_{j} | > u\Big)
\leq 
\frac{m}{(u k)^{\theta}} \Bigg( c_{\theta} \sum_{j=0}^{\infty}  \Theta_{j,{\theta}}^2 \Bigg)^{\theta/2},
\]
where $\sum_{j=0}^{\infty}  \Theta_{j,{\theta}}^2 < \infty$ due to Condition~3.1. 
In other words:
\[
\IP \Big(\max_{j=1,\ldots,m} | R_{j} - \tilde{R}_{j} | > M \frac{m^{1/\theta}}{k} \Big) = \mathcal{O}( M^{-{\theta}} ),
\]
will be arbitrarily small for $M$ large enough.
This concludes the proof of Lemma~\ref{lem:rate_Delta_s}. \hfill\qedsymbol

\noindent
\subsubsection{Proof of Lemma~\ref{lem:sum_Ber}}
\label{sec:proofs:sum_Ber}

Note that $S_n = \sum_{i=1}^n (2 I_i - 1) =: \sum_{i=1}^n E_i$ is a random walk that starts in $S_0 = 0$. The first time $n>0$ that yields $S_n = 0$ again is $t_0 := \inf\{n > 0 : S_n = 0\}$ and we have
\begin{equation*}
\begin{split}
\IP\Big( \max_{j \geq 1} \sum_{i=1}^j (2 I_i - 1) > 0 \Big)
& = \sum_{k=1}^{\infty} \IP(t_0 = 2k)
\end{split}
\end{equation*}
We have, for $k>0$, that
\begin{equation*}
\begin{split}
\IP(t_0 = 2k)
& = \sum_{(i_1, \ldots, i_{2k}) \in A_k} \IP(E_1 = i_1, \ldots, E_{2k} = i_{2k}). \\ 
\end{split}
\end{equation*}
The assertion follows from the assumed independence of $I_1, I_2, \ldots$. \hfill\qed

\noindent
\subsubsection{Proof of Lemma~\ref{lem:mu1hat}}
\label{sec:proofs:mu1hat}

For the proof of~\eqref{eqn:rate_mu1hat} note that
\begin{equation*}
\begin{split}
    \IP\Big( | \hat\mu_1 - \mu_1 | > \varepsilon \Big)
    & \leq \IP\Big( \Big| \frac{1}{k \hat\eta} \sum_{i = 1}^{k \hat\eta} (X_i - \mu_1) \Big| > \varepsilon, \hat\eta = \eta \Big)
    + \IP\Big( \hat\eta \neq \eta \Big) \\
    & \leq \IP\Big( \Big| \sum_{i = 1}^{k \eta} Z_i \Big| > \varepsilon k \eta \Big) + o(1) \\
    & \leq C_1 \frac{k \eta \Xi_{\theta, 1}^{\theta}}{( \varepsilon k \eta )^{\theta}} + C_2 \exp\Big(- \frac{C_3 (\varepsilon k \eta)^2}{ k \eta \Xi_{2, 1}^{2} } \Big)  + o(1) \\
    & = C_1 \Xi_{\theta, 1}^{\theta} \Big( \frac{ 1 }{ \varepsilon (k \eta)^{(\theta-1)/\theta} } \Big)^{\theta} + C_2 \exp\Big(- \frac{C_3}{\Xi_{2, 1}^{2}} \big( \varepsilon (k \eta)^{1/2} \big)^2 \Big)  + o(1), \\
\end{split}
\end{equation*}
which implies~\eqref{eqn:rate_mu1hat}, since $(k \eta)^{(\theta-1)/\theta} = (k \eta)^{1/2} (k \eta)^{(\theta-2)/(2\theta)}$ with $(\theta-2)/(2\theta) > 0$.\hfill\qed

\noindent
\subsubsection{Proof of Lemma~\ref{lem:dhat}}
\label{sec:proofs:dhat}

Further, for the proof of~\eqref{eqn:rate_dhat}, recall
\[\hat{d} : = \min_{\substack{i =  k (\hat{\eta} + 1) + 1,\\ \ldots, n-k+1}} \ \frac{1}{k} \sum_{j = i}^{i + k-1} (X_j - \hat{\mu}_1)\]
and $d_*$ defined in~(20) as
\begin{equation*}
d_* : = \min_{\substack{i =  k (\eta + 1) + 1,\\ \ldots, n-k+1}} \ \frac{1}{k} \sum_{j = i}^{i+ k-1} (\mu_j - \mu_1).
\end{equation*}

Due to $| \min_i x_i - \min_i y_i| \leq \max_i | x_i - y_i |$ (in the second inequality), we have
\begin{align*}
    & \IP\Big( | \hat d - d_* | > \varepsilon \Big)
    \leq \IP\Big( | \hat d - d_* | > \varepsilon, \hat\eta = \eta \Big) + \IP\Big( \hat\eta \neq \eta \Big) \\
    & \leq \IP\Big( \min_{\substack{i =  k (\eta + 1) + 1,\\ \ldots, n-k+1}} \ \Big| \frac{1}{k} \sum_{j = i}^{i + k-1} (X_j - \mu_j + \mu_1 - \hat\mu_1) \Big| > \varepsilon \Big) + o(1)\\
    & \leq \IP\Big( \min_{\substack{i =  k (\eta + 1) + 1,\\ \ldots, n-k+1}} \Big| \frac{1}{k} \sum_{j = i}^{i + k-1} Z_j \Big| > \varepsilon/2, | \hat \mu_1 - \mu_1| \leq \varepsilon/2 \Big) + \IP\Big( | \hat \mu_1 - \mu_1 | > \varepsilon/2 \Big) + o(1)\\
    & \leq (m - \eta + 1) \IP\Big( \max_{i = 1, \ldots, 2 k} \Big| \sum_{j = 1}^{i} Z_j \Big| > k \varepsilon/4 \Big) + \IP\Big( | \hat \mu_1 - \mu_1 | > \varepsilon/2 \Big) + o(1)\\
    & \leq (m - \eta + 1)  C_1 \frac{2 k \Xi_{\theta, 1}^{\theta}}{(k \varepsilon/4)^{\theta}} + (m - \eta + 1) C_2 \exp\Big(- \frac{C_3 (k \varepsilon/4)^2}{2k \Xi_{2, 1}^{2} } \Big) \\
    & \qquad +\IP\Big( | \hat \mu_1 - \mu_1 | > \varepsilon/2 \Big) + o(1) \\
    & \leq C_1 \frac{2 \Xi_{\theta, 1}^{\theta}}{( ((m - \eta + 1) k^{1-\theta} )^{-1/\theta} \varepsilon/4)^{\theta}} + C_2 \exp\Big( \log( m - \eta + 1 ) - k \varepsilon^2 \frac{C_3}{32\Xi_{2, 1}^{2}} \Big) \\
    & \qquad + \IP\Big( | \hat \mu_1 - \mu_1 | > \varepsilon/2 \Big) + o(1).
\end{align*}

Choose $\varepsilon_M := M \max\{ ((m - \eta + 1) k^{1-\theta})^{1/\theta}, \tilde C (\log(m - \eta + 1)/k)^{1/2} \}$,\\
for some $\tilde C \geq \big( 64 \Xi_{2, 1}^{2} /C_3 \big)^{1/2}$.
By assumption, we have $n-\tau \geq 2 k$ such that $m-\eta+1 \geq 2$.
Thus,
\begin{equation*}
    \begin{split}
        \IP\Big( | \hat d - d_* | > \varepsilon_M \Big)
        & \leq C_1 \frac{2 \Xi_{\theta, 1}^{\theta}}{( M/4 )^{\theta}} + C_2 \exp\Big( - M^2 \log(2) \Big)  \\
        & \quad + \IP\Big( | \hat \mu_1 - \mu_1 | > M \tilde C (k\eta)^{-1/2} / 2 \Big) + o(1),
    \end{split}
\end{equation*}
where in the second line we have used 
the fact that $(k\eta)^{-1/2} \leq (\log(m-\eta+1)/k)^{1/2}$.
Recalling Lemma~\ref{lem:mu1hat}, we can thus choose $M$ large enough for the bound to be arbitrarily small.
Therefore, we have proved
\[\hat d = d_* + \mathcal{O}_{\IP}( \max\{(\log(m-\eta+1)/k)^{1/2}, (m k^{1-\theta})^{1/\theta}\}).\]

\enlargethispage{1cm}
Finally, by condition (C1), we have
\[ m^{1/\theta} k^{(1-\theta)/\theta} = \frac{(m k)^{1/\theta}}{k^{1/2} k^{1/2}}\leq \frac{1}{k^{1/2}} \frac{n^{1/\theta}}{\sqrt{k}} = \frac{1}{k^{1/2}} \mathcal{O}( 1 ), \]
which finishes the proof.
\hfill\qed

\clearpage
\section{Further results from simulation experiments}\label{app:FurtherSim}

In Table~\ref{tab:ratio_null_lrv_est} empirical sizes are shown for a simulation experiment that is identical to the one described in~Section~4.2, but for the fact that the long-run variance is estimated by $\hat\sigma^2$, defined in~(8), using the general definition of $\hat L$ and $J=3$.

It can be seen that the type-I error rates are sometimes higher than the nominal level of $\alpha = 0.05$, especially for small sample sizes and when the dependence is strong. As $n$ increases the rejection ratios approach the nominal level, as expected.

\begin{figure}[t]
	\centering
	\includegraphics[width=\textwidth]{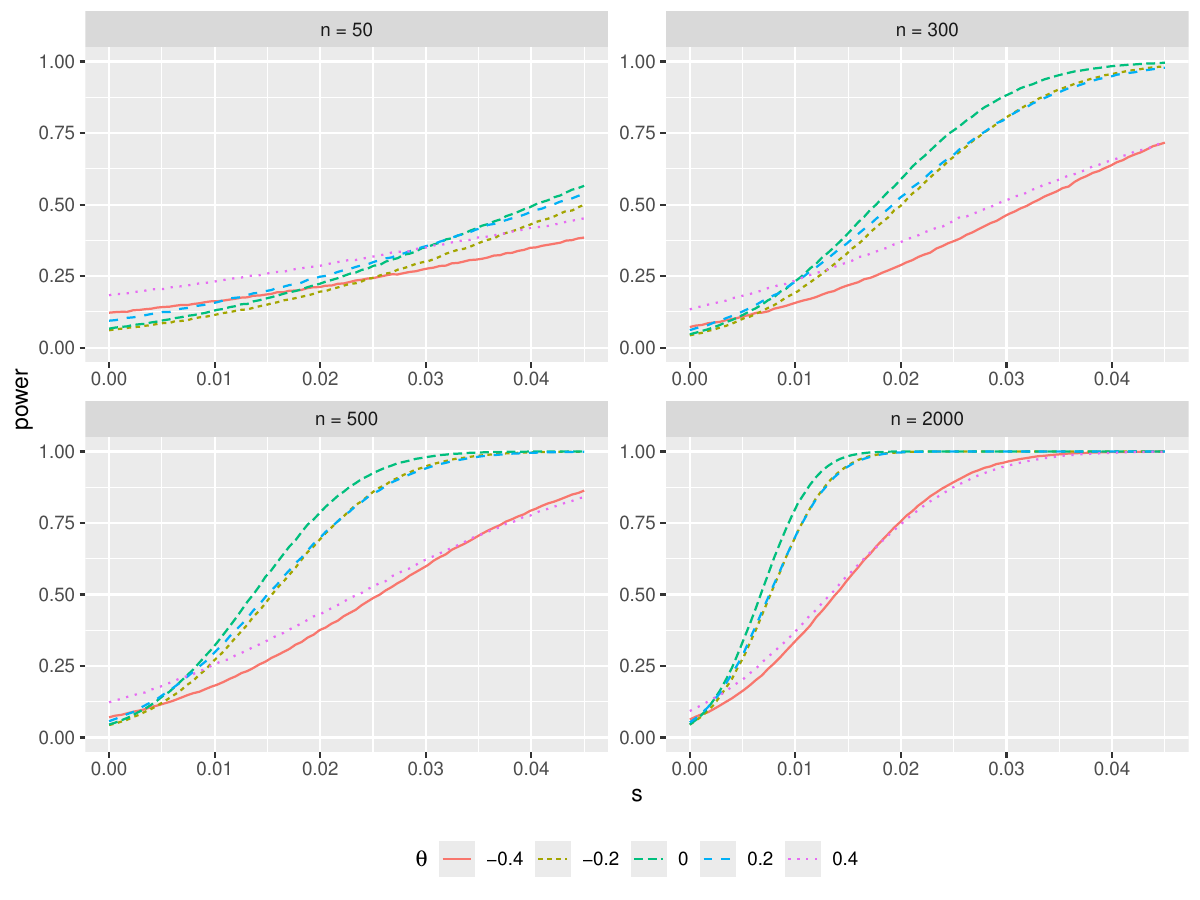}
	\caption{Rejection ratios for the testing procedure using $\hat\sigma$ and asymptotic quantile, under the alternative hypothesis: \(\mu_i = \mu_1\) for \(i = 1,2,\ldots,\tau-1\); \(\mu_i > \mu_1 + s\) for \(i = \tau, \tau + 1, \tau + 2, \ldots,n\). The noise process is captured by the dependence parameter \(\theta\). We adjust the gap parameter \(s\) over the set \(\{0, 0.0006, 0.0012, \ldots, 0.045\}\), \(n\) over \(\{50, 300, 500, 2000\}\), and \(\theta\) over \(\{-0.4, -0.2, 0, 0.2, 0.4\}\). Each data point represents $100,000$ replications.}
	\label{Fig:ratio_alt_lrv_est}
\end{figure}

The power of the test with long-run variance estimated by $\hat\sigma^2$ is illustrated in Figure~\ref{Fig:ratio_alt_lrv_est}.

\begin{figure}[b]
\centering
\includegraphics[width=\textwidth]{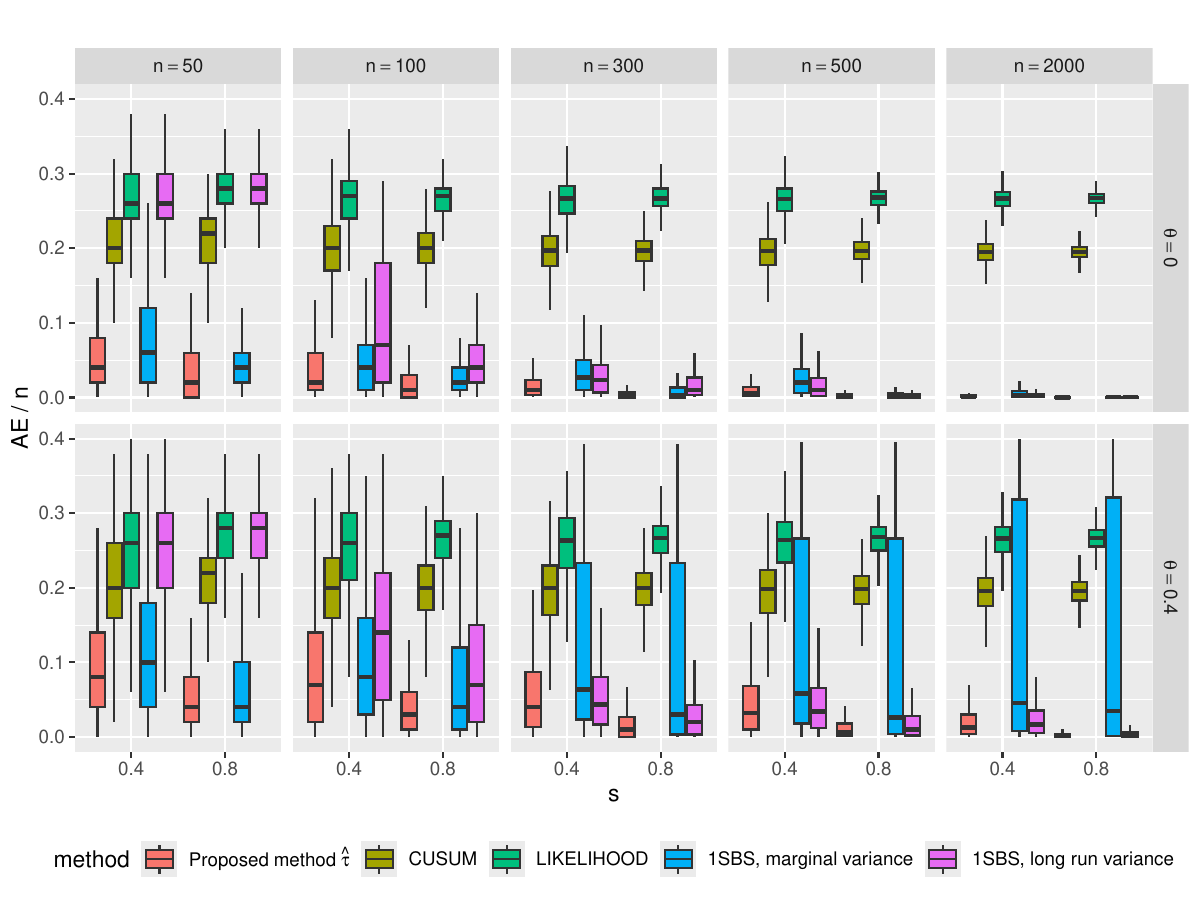}
\caption{\label{Fig:box_plot_five_methods}  Box plots of scaled absolute error (MAE/$n$) comparing our method, the CUSUM method,
likelihood-based method, binary segmentation using marginal variance, and binary segmentation using
long-run variance.
}
\end{figure}

We provide box plots of the simulated rescaled absolute error rate of five different methods and summarize the results in Figure~\ref{Fig:box_plot_five_methods}. The simulation setup is exactly the same as that in Figure~5.

\begin{table}[t]
    \caption{\label{tab:ratio_null_lrv_est}Rejection ratios for change point testing procedure under the null hypothesis using $\hat \sigma$.}
    {\footnotesize
        \hspace*{-0.2cm}
        \begin{tabular}{ccccccc} 
            \toprule 
            Approximation & & \multicolumn{4}{c}{$\theta$} \\ 
            \cmidrule{3-7} 
             Method & $n$ & {-0.4} & {-0.2} & {0}  & {0.2}  & {0.4} \\ 
            \midrule 
            Asymptotic 
            &    50 & 12.2\% & 6.09\% & 6.65\% & 9.44\% & 18.3\% \\
            &   100 & 9.38\% & 4.77\% & 5.27\% & 7.52\% & 15.7\% \\
            &   300 & 7.22\% & 4.31\% & 4.71\% & 6.14\% & 13.4\% \\
            &   500 & 7.08\% & 4.40\% & 4.53\% & 5.72\% & 12.3\% \\
            &  2000 & 6.32\% & 4.52\% & 4.53\% & 5.32\% & 9.27\% \\
            \addlinespace
            Finite-sample
            &    50 & 14.5\% & 7.86\% & 8.49\% & 11.5\% & 20.5\% \\
            &   100 & 11.2\% & 6.11\% & 6.54\% & 9.02\% & 17.5\% \\
            &   300 & 8.30\% & 5.06\% & 5.45\% & 7.03\% & 14.5\% \\
            &   500 & 7.96\% & 4.98\% & 5.11\% & 6.40\% & 13.2\% \\
            &  2000 & 6.70\% & 4.80\% & 4.83\% & 5.65\% & 9.68\% \\
            \bottomrule 
    \end{tabular} 
}
\end{table}


\bibliography{references.bib}